\documentclass[10pt,conference]{IEEEtran}
\IEEEoverridecommandlockouts
\usepackage{cite}
\usepackage{amsmath,amssymb,amsfonts}

\usepackage[ruled,vlined, linesnumbered]{algorithm2e}
\SetKwInput{KwInput}{Input}                
\SetKwInput{KwOutput}{Output}              
\SetKwRepeat{Do}{do}{while}
\usepackage{xcolor}

\SetCommentSty{mycommfont}
\usepackage[htt]{hyphenat}
\usepackage[hyphens]{url}
\usepackage{lstautogobble}
\UseRawInputEncoding

\usepackage{array}
\usepackage{multicol}
\usepackage{multirow} 
\usepackage{caption}
\usepackage{physics}
\usepackage{array} 
\usepackage{makecell} 
\usepackage{subcaption}
\usepackage{graphicx}
\usepackage{textcomp}
\usepackage{xcolor}
\usepackage{fontawesome}
\usepackage{tikz}
\newcommand*\circled[1]{\tikz[baseline=(char.base)]{
            \node[shape=circle,draw,inner sep=0.75pt, text=white,fill=black] (char) {#1};}}
\usepackage{listings}
\colorlet{punct}{red!60!black}
\definecolor{background}{HTML}{FFFFFF}
\definecolor{delim}{RGB}{20,105,176}
\colorlet{numb}{magenta!60!black}
\lstdefinelanguage{json}{
    basicstyle=\footnotesize\ttfamily,
    numbers=left,
    numberstyle=\scriptsize,
    stepnumber=1,
    numbersep=1pt,
    showstringspaces=false,
    breaklines=true,
    frame=lines,
    backgroundcolor=\color{background},
    literate=
     *{0}{{{\color{numb}0}}}{1}
      {1}{{{\color{numb}1}}}{1}
      {2}{{{\color{numb}2}}}{1}
      {3}{{{\color{numb}3}}}{1}
      {4}{{{\color{numb}4}}}{1}
      {5}{{{\color{numb}5}}}{1}
      {6}{{{\color{numb}6}}}{1}
      {7}{{{\color{numb}7}}}{1}
      {8}{{{\color{numb}8}}}{1}
      {9}{{{\color{numb}9}}}{1}
      {:}{{{\color{punct}{:}}}}{1}
      {,}{{{\color{punct}{,}}}}{1}
      {\{}{{{\color{delim}{\{}}}}{1}
      {\}}{{{\color{delim}{\}}}}}{1}
      {[}{{{\color{delim}{[}}}}{1}
      {]}{{{\color{delim}{]}}}}{1},
}
\newcolumntype{P}[1]{>{\centering\arraybackslash}p{#1}}
\usepackage[nolist]{acronym}

\usepackage{float}

\begin{acronym}

\acro{faas}[FaaS]{Function-as-a-Service}
\acro{iaas}[IaaS]{Infrastructure-as-a-Service}
\acro{paas}[PaaS]{Platform-as-a-Service}
\acro{baas}[BaaS]{Backend-as-a-Service}
\acro{name}[SLAM]{SLAM}
\acro{fdn}[FDN]{Function Delivery Network}
\acro{hpc}[HPC]{High-Performance Computing}
\acro{slo}[SLO]{Service-Level Objective}
\acro{sla}[SLA]{Service-Level Agreements}
\acro{rtt}[RTT]{Round Trip Time}
\acro{vu}[VU]{Virtual User}
\acro{rt}[RT]{Response Time}
\acro{mms}[MMS]{Minimum Memory Sum}
\acro{moc}[MOC]{Minimum Overall Cost}
\acro{csp}[CSP]{Cloud Service Provider}
\acro{moet}[MOET]{Minimum Overall Execution Time}

\end{acronym}
\def\BibTeX{{\rm B\kern-.05em{\sc i\kern-.025em b}\kern-.08em
    T\kern-.1667em\lower.7ex\hbox{E}\kern-.125emX}}

\begin{document}

\title{SLAM: SLO-Aware Memory Optimization for Serverless Applications\\
{\footnotesize \textsuperscript{*}Note: This the preprint version of the accepted paper at IEEE CLOUD'22}
}

\author{\IEEEauthorblockN{Gor Safaryan, Anshul Jindal, Mohak Chadha, Michael Gerndt\\}
\IEEEauthorblockA{Chair of Computer Architecture and Parallel Systems, Technische Universit{\"a}t M{\"u}nchen, Germany \\ Garching (near Munich), Germany \\
Email: \{gor.safaryan, anshul.jindal, mohak.chadha\}@tum.de, gerndt@in.tum.de}}

\maketitle

\begin{abstract}
Serverless computing paradigm has become more ingrained into the industry, as it offers a cheap alternative for application development and deployment. This new paradigm has also created new kinds of problems for the developer, who needs to tune memory configurations for balancing cost and performance. Many researchers have addressed the issue of minimizing cost and meeting Service Level Objective (SLO) requirements for a single FaaS function, but there has been a gap for solving the same problem for an application consisting of many FaaS functions, creating complex application workflows.

In this work, we designed a tool called SLAM to address the issue. SLAM uses distributed tracing to detect the relationship among the FaaS functions within a serverless application. By modeling each of them, it estimates the execution time for the application at different memory configurations. Using these estimations, SLAM  determines the optimal memory configuration for the given serverless application based on the specified SLO requirements and user-specified objectives (minimum cost or minimum execution time). We demonstrate the functionality of SLAM on AWS Lambda by testing on four applications. Our results show that the suggested memory configurations guarantee that more than 95\% of requests are completed within the predefined SLOs.

\end{abstract}


\section{Introduction}
Significant progress has been made in different domains~\cite{grafberger_fedless2021,carreira2019cirrus, shankar2018numpywren, jonas2017occupy, Fox2017} based on the idea of \textit{serverless computing} since its launch by Amazon as AWS Lambda in November 2014~\cite{awslambdarelease}. Serverless computing is a cloud computing model that abstracts server management and infrastructure decisions away from the users~\cite{wg2018cncf}. In this model, the allocation of resources is managed by the cloud service provider rather than by \textit{DevOps}, thereby benefiting them from various aspects such as no infrastructure management, automatic scalability, and faster deployments~\cite{cloudflareWhyUse, Roberts2018}.  Function-as-a-Service (FaaS) is a key enabler of serverless computing~\cite{wg2018cncf}. In FaaS, a serverless application is decomposed into simple, standalone functions that are uploaded to a FaaS platform such as AWS Lambda~\cite{AWSLambda:online}, Google Cloud Function (GCF)~\cite{GoogleCloudFunctions:online}, and Azure Functions (AF)~\cite{AzureFunctions:online} for execution. Most of the public cloud providers in their FaaS offerings allow users to configure memory allocation for the functions~\cite{AWSLambda, GoogleCloudFunctions:online, Azure}.

\begin{figure*}[t]
\centering
\begin{subfigure}{0.27\textwidth}
   \centering
    \includegraphics[width=1\linewidth]{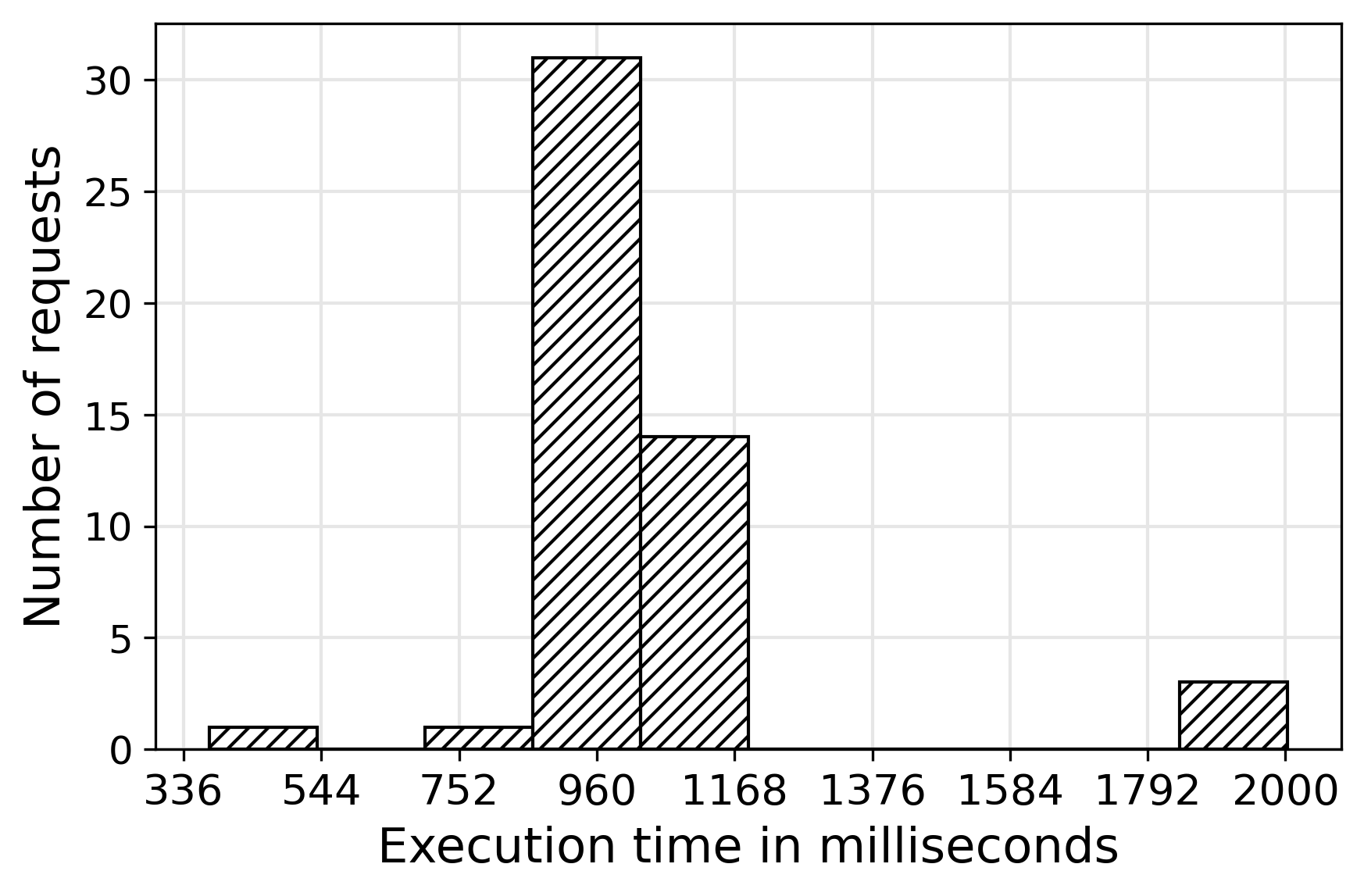}
    \caption{Execution time variance due to cold start problem.} \label{fig:dist128}
\end{subfigure} \hspace{2pt}
\begin{subfigure}{0.27\textwidth}
    \centering
    \includegraphics[width=1\linewidth]{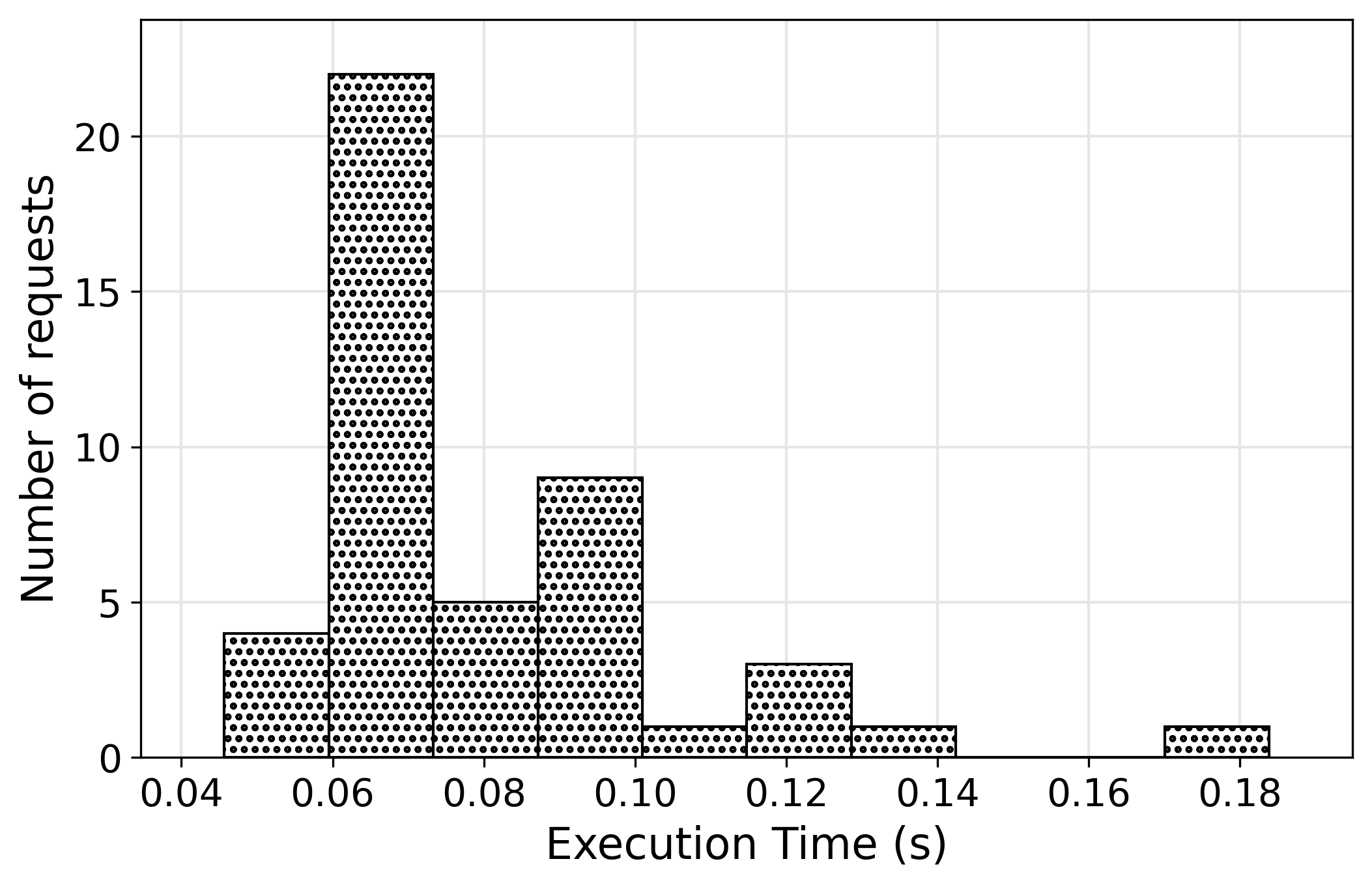}
    \caption{Execution time variance due to BaaS service (DynamoDB). } \label{fig:dist128Baas}
\end{subfigure} \hspace{2pt}
\begin{subfigure}{0.27\textwidth}
   \centering
    \includegraphics[width=1\linewidth]{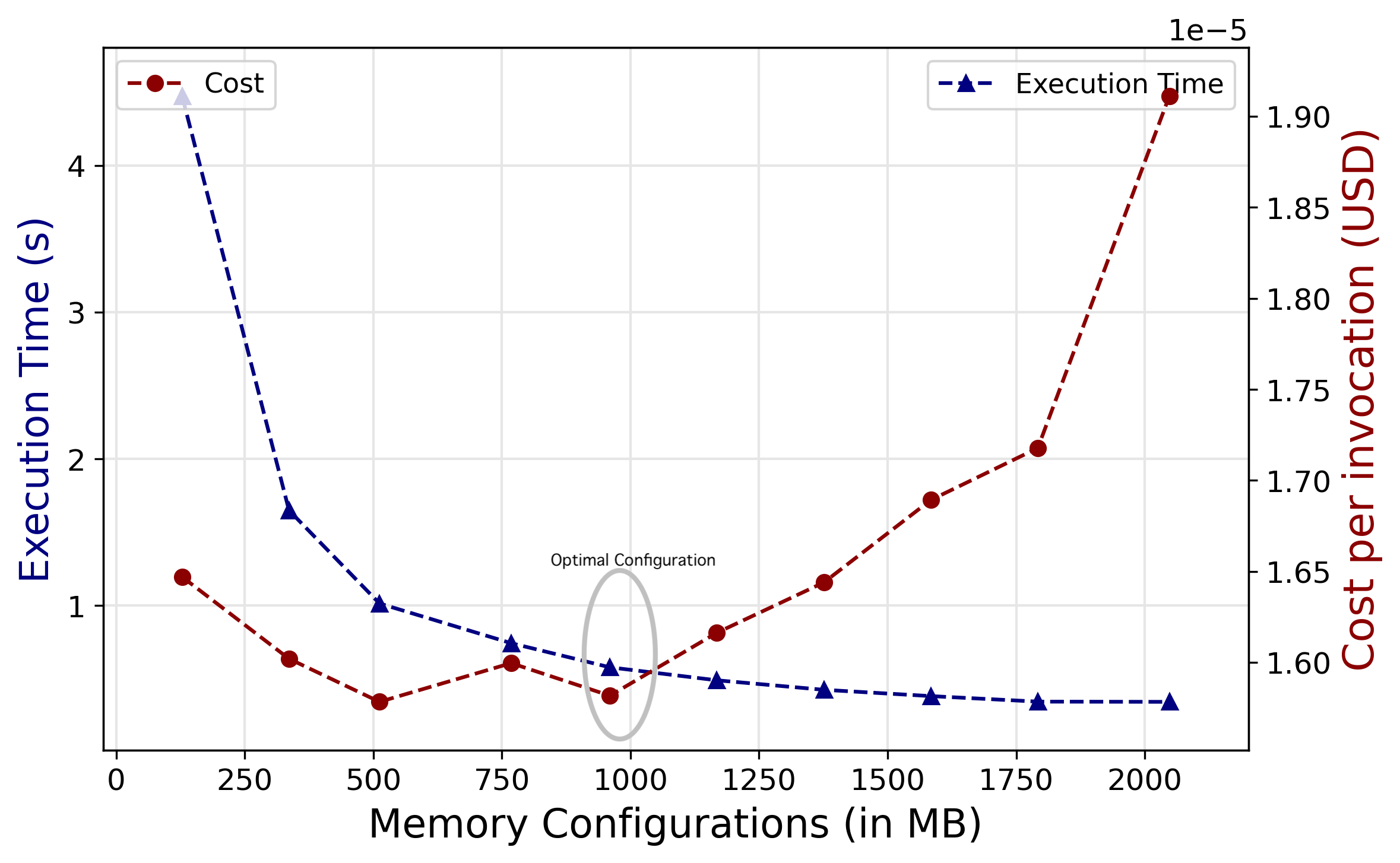}
    \caption{Performance vs cost trade for finding the optimal configuration.} \label{fig:optimal_config_perf_cost} 
\end{subfigure}
\caption{
Various factors making it difficult to optimally configure the memory of the FaaS functions within a serverless application.}
\label{fig:slo_defining_factors}
\end{figure*}

Despite having many advantages, serverless computing suffers from some pain points that obstruct its wide adoption~\cite{Baldini2017,fdn_jindal2021, DBLP:journals/cloudcomp/Eivy17}. The most commonly known is optimally configuring the memory of the FaaS functions within the application based on the required the Service Level Objective (SLO). The difficulties lie in the following aspects:\\
 \textbf{Cold start}: It is mainly connected with loading the FaaS function into the main memory of the executing server and preparing the execution environment for the target code (starting up the VM/container, loading libraries, loading function code, etc.)~\cite{mohan2019agile, Byrro2019}. The cold start phenomenon combined with the heterogeneity of the cloud environment makes the function execution time quite unpredictable. Figure~\ref{fig:dist128} shows an execution time distribution for a sample compute-intensive function having a high variance when deployed with $128$MB memory configuration on AWS Lambda. \\
 \textbf{FaaS functions integration with BaaS services}:  The FaaS functions are usually closely integrated with other services, e.g., cloud databases, authentication and authorization services, and messaging services. These services are called Backend-as-a-Service (BaaS)~\cite{lane2015overview}. These services also do influence the execution time of the FaaS functions, thus adding the variance in the time. Figure~\ref{fig:dist128Baas} shows an execution time distribution for a sample function querying DynamoDB having a high variance when deployed with $128$MB memory configuration on AWS Lambda. \\
 \textbf{Trade-off analysis between performance and cost}: Users need to define memory configuration for their FaaS functions: a low-level information which directly influences the performance and cost of the serverless application~\cite{AWSLambda:online, DBLP:conf/middleware/Spillner20, wang2018peeking}. Thus, the user has to do a trade-off analysis between them to define the right configuration for their required SLOs~\cite{specResearchVisionFaas}. Figure~\ref{fig:optimal_config_perf_cost} shows an execution time vs the cost graph for a sample compute-intensive function when deployed with different memory configurations on AWS Lambda. We can observe that it's not trivial to find the optimal configuration where the overall cost and execution time are both optimal. \\
 \textbf{Complex application workflows}: Usually, the serverless applications comprise dozens if not hundreds of small FaaS functions and these connect together to form complex event-driven workflows. Furthermore, the SLOs are usually defined at the application level instead of the function level and thus based on the required application SLOs configuring the memory of the FaaS functions within the application even becomes more challenging since a change in one can influence the others.


The aspects above highlight some factors that make it difficult for the users to optimally configure memory for serverless applications based on the required SLOs. However, there are many other factors such as I/O and network bandwidth, and co-location with other functions affecting the performance and cost which the users are not aware of~\cite{wang2018peeking}. Many researchers have addressed the issue of optimizing the memory and cost for meeting SLO requirements for a single cloud function~\cite{cose, costless, eismann2021sizeless}, but there has been a gap for solving the same problem for a serverless application consisting of many FaaS functions, which create a complicated workflow of function calls. To this end, we develop \textbf{SLAM}, a python-based tool that can automatically find the optimal memory configurations for the FaaS functions within the given serverless application based on the specified SLO. Our key contributions are as follows: 
\begin{itemize}
        \item We develop and present a novel tool called \textbf{SLAM} that automatically determines the optimal memory configuration for the FaaS functions within the given serverless application based on the specified SLO requirements (\S\ref{sec:system_description}). To the best of our knowledge, this is the first work that find and configure FaaS functions with optimal memory configurations within a serverless application based on the specified SLO.

        \item We propose and implement an optimization algorithm along with its variants for various optimization objectives (minimum cost and minimum overall time) in addition to the SLO requirements in finding the optimal memory configuration for the given serverless application (\S\ref{sec:algorithms}).
        
        \item Although our approach is generic and \textit{SLAM} can be easily extended to support other commercial and open-source FaaS platforms, we demonstrate the functionality of \textit{SLAM} with AWS Lambda (\S\ref{sec:evaluation}) on four serverless applications comprising of various number of functions. 
                
        \item We evaluate the performance of the \textit{SLAM} on $3$ different aspects: 1) Estimation time accuracy (\S\ref{sec:est_time_accuracy}), 2) Configuration finding accuracy (\S\ref{sec:config_finding_accuracy}), and 3) Configuration finding efficiency and scalability of \textit{SLAM} (\S\ref{sec:config_finding_efficency_scalability}). From the experimental evaluation, the suggested memory configurations guarantee that more than 95\% of requests are completed within the defined SLOs.
\end{itemize}
\begin{table}[t]
\centering
\caption{Symbols and definitions used in this paper}\label{tab1:symbols}
\begin{tabular}{m{2.5cm}|m{5.5cm}}
\hline
\textbf{Symbol}&\textbf{Interpretation}\\
\hline
$N$ & total number of functions in an application \\

$M$ & total number of memory configurations \\
$S$ &  total number of sequence groups formed from application call graph \\
$U_i$ &  total number of sub-sequence groups within some group $i$ \\
$K$ &  total number of user-requests for load generation \\
$X$ & possible number of memory configurations adhering to the defined SLOs.\\
$m_i^j$ & memory allocated to $i^{th}$ function in the $j^{th}$ configuration set \\
$mem\_config\_list$ & a  list  of  memory values [128, 256, 512, 1024, 2048, 4096, 8192, 10240]   \\

$F = \{f_1, \dots,  f_N$\} & functions within an application \\
$G = \{g_1, \dots, g_{S} \}$ & sequence groups from application's call graph\\
$\bar{G} = \{\bar{g}_1^i, \dots, \bar{g}_U^i \} $ & sub-sequence groups within some group $i$ \\
$C = \{C_1, \dots, C_X \}$ & memory configs adhering to the defined SLOs.\\

$\alpha$ & $n^{th}$ percentile (called choice percentile) of the distribution as a representative for the execution time for the given function at a particular memory configuration. \\

\hline
\end{tabular}
\end{table}

\section{System Description}
\label{sec:system_description}
In this section,  we present \textit{SLAM}, a python-based tool for automatically configuring the FaaS functions within a serverless application with optimal memory such that the overall execution time of invocations to the application conform to the defined \ac{slo} requirements. In this work, we consider the $95^{th}$ percentile execution time of an application invocation as the SLO. \textit{SLAM} also supports additional user-specified objectives on top of the \ac{slo} requirements: 1) \ac{moc}, and 2) \ac{moet}, by which the \textit{SLAM} suggested configuration for the serverless application not only conforms to the defined \ac{slo} requirements, but also meets user-specified objectives. \textit{SLAM} can dynamically adapt to changes in the given serverless application and automatically adjust memory configurations of functions. \textit{SLAM} can be incorporated into a \ac{csp} \ac{faas} platform and then leveraged by application developers for optimizing the memory configuration of their serverless applications.

\begin{figure}[t]
    \centering
    \includegraphics[width=0.9\linewidth]{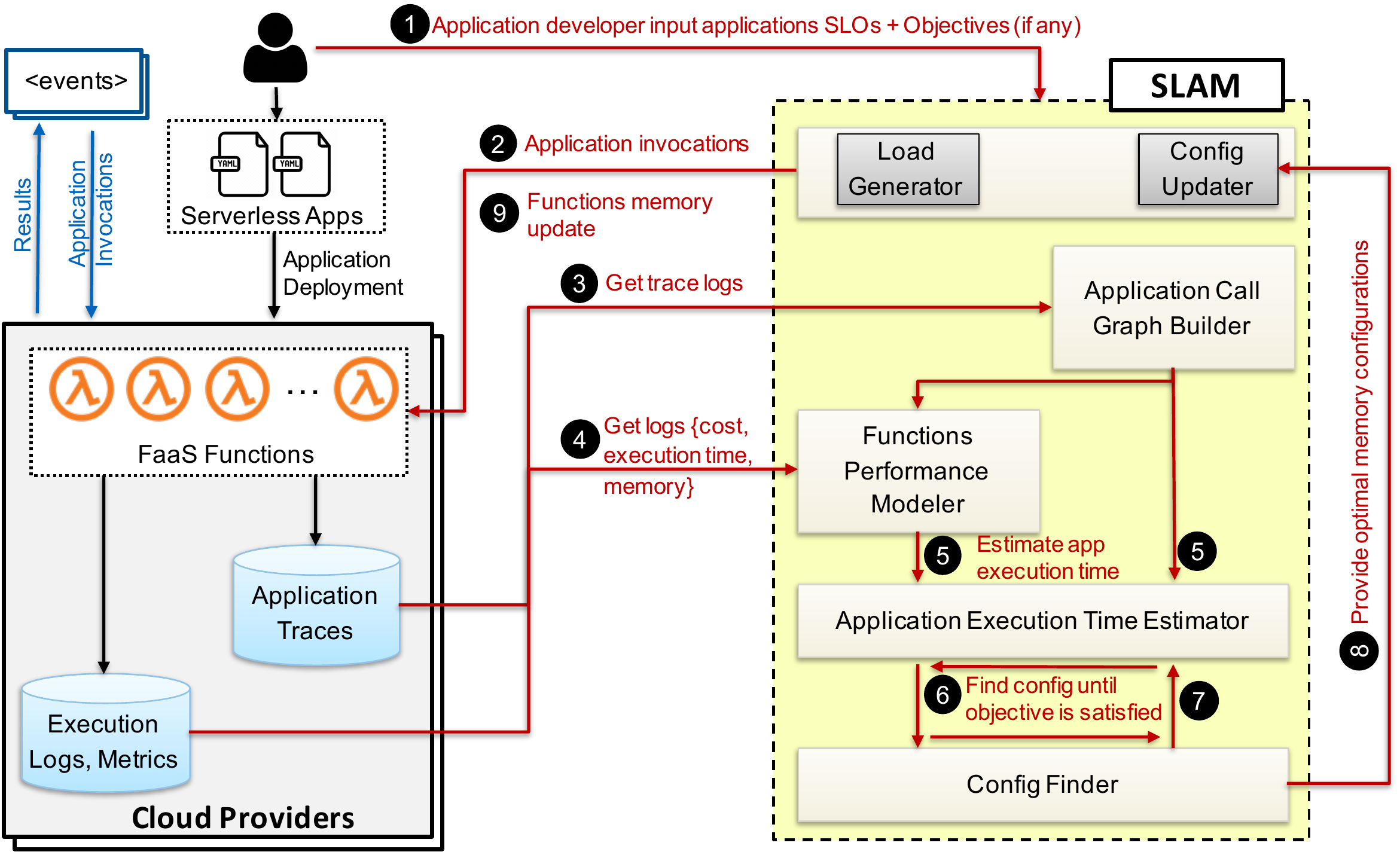}
    \caption{High-level architecture of the \textit{SLAM} and the interaction between its components in a general use case.} \label{fig:overall_architecture}
\end{figure}


Figure~\ref{fig:overall_architecture} provides an overview of our developed \textit{SLAM} tool and the interaction between its components in a typical usecase. \textit{SLAM} assumes that the serverless application which is to be configured is already deployed by the application developer on a \ac{faas} platform (AWS Lambda~\cite{AWSLambda:online} in our case) and is instrumented with a middleware tracing library (such as AWS X-Ray~\cite{awsXRAY}) to trace the incoming and outgoing requests to other functions, or other cloud components/services. 

\textit{SLAM} takes the \acp{slo} requirement for the application as the input along with other user-specified objectives (if any) (step \circled{1}). Then the \textit{Load Generator} component of it generates a minimal amount of user workload ($K=50$ application invocations, see Table~\ref{tab1:symbols}) to the application's public endpoint (step \circled{2}) and collects application trace logs (step \circled{3}) and various monitoring metrics\footnote{https://docs.aws.amazon.com/lambda/latest/dg/monitoring-metrics.html} data (step \circled{4}). The collected logs are used by \textit{Application Call Graph Builder} component to construct the application call graph (step \circled{3}) and this call graph along with the monitoring metrics are further used by \textit{Functions Performance Modeler} component for building the application's functions performance models. \textit{Application Execution Time Estimator} component use models along with the application call graph for estimating the overall application response time on the different memory configurations provided by \textit{Config Finder} component. \textit{Config Finder} component generates the configuration based on the developed algorithms (\S \ref{sec:algorithms}) and examine the estimated time, memory configurations, and cost for the \acp{slo} requirements and user-specified objective (if any) satisfaction (step \circled{6}). If the \acp{slo} requirements and user-specified objective are not satisfied, \textit{Config Finder} tries different memory configurations (step \circled{7}) and continues the process until it is satisfied (steps \circled{6} - \circled{7}). Once a configuration is found, the functions' memory configurations are updated by \textit{Config Updater} component (steps \circled{8} - \circled{9}). Next, we describe the six major components of \textit{SLAM} tool in more detail.

\subsection{Load Generator}
This component is responsible for generating user workload to the deployed application. It takes a total number of requests to the application as input and based on it generates the given amount of user workload requests synchronously to the deployed application. This user workload generation allows creating application traces and collect various metrics data used by the other components of the \textit{SLAM}.

\subsection{Application Call Graph Builder} 
\label{sec:call_graph_builder}
This component is responsible for building the application call graph involving the application functions and \ac{baas} services such as database, storage, and queues. \textit{SLAM} relies on external middleware tracing libraries (such as AWS X-Ray) instrumented by the application developer allowing to trace the incoming and outgoing requests to other functions, or \ac{baas} services. 

This component with the help of  \textit{Load Generator} component generates a small amount of user workload requests (\S \ref{sec:evaluation_settings}) to the deployed application. The application traces are then parsed to generate the application call graph involving all the functions and \ac{baas} services within the application. Afterward, the component filters out \ac{baas} services, as it is out of the scope of this work to tune them. 
As a result, after this step, we get the simplified call graph for the deployed serverless application along with the composing functions. In case the user already has the application call graph and wants to skip this step, \textit{SLAM} allows the user to input manually the call graph of the application. This also increases the testability of the \textit{SLAM} for further development.

\subsection{Functions Performance Modeler}
\label{sec:functions_performance_modeling}
After building the call graph of the application and knowing its composing functions, the next step is to estimate the execution time of each function within the serverless application at different memory configurations. This is done in two steps, explained next.

\subsubsection{Create traces and metrics data for building models}
This component with the help of  \textit{Load Generator} component first generates a small amount of user workload requests ($K=50$ application invocations, see Table~\ref{tab1:symbols}) to the deployed application when all of its composing functions are deployed with a default same memory configuration ($128$MB). Based on the composing functions found by the \textit{Application Call Graph Builder} component, it then requests \textit{Config Updater} component for updating the memory configurations of those functions based on the default list of memory configuration values ($mem\_config\_list$ in Table~\ref{tab1:symbols}) and \textit{Load Generator} to again generate the same amount of user workload requests to the updated application.  The process is repeated for all the memory configurations ($mem\_config\_list$ in Table~\ref{tab1:symbols}) and in the end application traces and various metrics data are created for estimating  execution time for each function within the application.

\subsubsection{Estimation of execution time for each function }
Traces are parsed and metrics are analyzed to create a distribution of execution time for each function and each memory configuration. An example of such a distribution for a test function, when deployed with 128MB memory configuration on AWS Lambda, is shown in Figure~\ref{fig:dist128}.

One can observe that there is a high variance in the execution time of the function running under the same configuration due to the uncertainties from the underneath virtualized cloud infrastructure such as co-location of functions, cold-start, hardware failures, resource-overuse, etc. Therefore,  to overcome this inherent variance, we choose a hyperparameter called \textit{choice percentile} ($\alpha$ in Table~\ref{tab1:symbols}) representing the $n^{th}$ percentile of the distribution as a representative for the execution time for the given function at a particular memory configuration. $\alpha$ is configured automatically by SLAM. Calculating prediction accuracy of execution time at multiple values of $\alpha$ (default test values: $50,75,90,99$), SLAM selects the one which results in a minimum mean squared error. 

Thus, in the end, a list of representative values for execution time for each function and memory combination is created, and how they are combined to form the overall execution time of the application is presented next. 

\subsection{Application Execution Time Estimator}
\label{sec:exec_tim_est}

Given the execution time of each FaaS function comprising the serverless application estimated by the \textit{Functions Performance Modeler} at certain memory configurations, it is the responsibility of this component to combine them to estimate the overall application execution time. 

Function invocations in the application can either be in parallel or one after the other in a sequence, or in a combination of both. Therefore, from the application call graph first, it determines which functions are executed in parallel to others by using the functions' start and end timestamps available from the traces. The tool then divides all functions into groups of sequence groups ($S$ in Table~\ref{tab1:symbols}), where all the functions in each group are executed in parallel to other functions in the same group, and each group is executed in sequence to other groups. Since all the functions in a group are invoked in parallel, therefore to estimate the execution time of a group we take the maximum of the execution times of all functions in the group. In the end, we sum the execution times of each group to get an estimate of overall application execution time. 

Mathematically, if we have an application consisting of $N$ functions configured with certain memory configurations and defined as $F = \{f_1, f_2, f_3, ...,  f_N$\}, with them being divided into $S$ sequence groups defined as $G = \{g_1, g_2, g_3, ..., g_{S} \}$, then the execution time of the whole application is given by: 

\begin{equation}
T(G) = \sum\limits_{x=1}^{N}F(g_x)\label{eq: duration}
\end{equation}
where for some group $i$: 
\begin{equation}
F(g_i) = \begin{cases}
    max(T(\bar{g}_1^i), \dots, T(\bar{g}_U^i)), & \text{if $g_i \neq$ function}.\\
    \text{function execution time},&\text{if $g_i =$ function}.
  \end{cases}
\end{equation}
where $\bar{g}_j^i$ ($1 \leq i \leq S$ and $1 \leq j \leq U$ ) being the sub-sequence group within $g_i$ and $U$ is the total number of sub-sequence groups within $g_i$. 

\subsection{Config Finder}
\label{sec:config_finder}
Given the estimated execution time for each FaaS function comprising the serverless application provided by the \textit{Functions Performance Modeler}, it is the responsibility of this component of \textit{SLAM} tool, \textit{Config Finder}, to find the right memory configurations for all functions such that the overall application execution time adheres to the defined SLOs and the specified optimization objectives (if any). We first present the two optimization objectives (\S\ref{sec:objectives_optimization}) that can be used as part of \textit{SLAM} tool in addition to the SLO requirements, and then we introduce the algorithm for finding the optimal memory configurations (\S\ref{sec:algorithms}).
\subsubsection{Optimization Objectives}
\label{sec:objectives_optimization}

Suppose there are total of $X$ possible memory configurations set for the serverless application defined as $C = \{C_1, C_2, ..., C_X \}$ such that  $C_j = \{m_1^j, m_2^j, ..., m_N^j\}$ ($1 \leq j\leq K $) is a memory configuration set for $F$ adhering to the defined SLOs and $m_i^j \in M$ ($1 \leq i\leq N $) is the memory allocated to $i^{th}$ function in the $j^{th}$ configuration set. 
Following are the two optimization objectives that can be used as part of \textit{SLAM} tool along with the defined SLOs: 

\textbf{Minimum Overall Cost (MOC)}: Here, the idea is to find a configuration that would result in a minimum cost for each invocation of the application under the given SLO requirements.  This is given by:

\begin{equation}
\begin{aligned}
\min_{j \in C} Cost(j) \label{eq:optimi_moc}
\end{aligned}
\end{equation}
where $Cost(j)$ ($j \in C$) is the overall application estimated cost when the application is configured with  $C_j$ configuration. Our calculation only counts for the costs associated with the function execution and does not take into account the data transfer, storage, and other costs associated with the invocation of functions. To calculate the aforementioned execution cost, we used the data provided by AWS~\cite{awslambdapricing}. Though they provide pricing only for a limited number of memory configurations, we interpolated the cost as there was a linear relationship between allocated memory and cost. 

\textbf{Minimum Overall Execution Time (MOET)}: The objective is to find a configuration that would result in minimum overall execution time of the application under the given SLO requirements. This  is then given by:
\begin{equation}
\begin{aligned}
\min_{j \in C} ExecTime(j) \label{eq: optimi_moet}
\end{aligned}
\end{equation}
where $ExecTime(j)$ ($j \in C$) is the overall application estimated time by \textit{Application Execution Time Estimator} when the application is configured with  $C_j$ configuration.

\subsubsection{Optimal memory configuration finding algorithm}
\label{sec:algorithms}
Now we describe the algorithm (called \textit{SLAM-SLO}) for finding the optimal memory configuration for serverless applications such that the overall application execution time adheres to the defined SLOs. The modified version of the algorithm for optimizing on various objectives along with the SLOs is called \textit{SLAM-SLO-Min-Cost} for \ac{moc} and \textit{SLAM-SLO-Min-Time} for \ac{moet}. We additionally compared our developed algorithm with brute force (referred as \textit{Brute-Force}) approach where all possible combinations for configurations for the functions within the application are generated to find the configuration that conforms to defined SLOs and the given objective. The overall complexity of this brute force approach is $O(M^N)$.

\textbf{SLAM-SLO}: In this approach, we leverage the max-heap data structure for finding the optimal configuration which satisfies the SLO requirements. The pseudocode for the algorithm is shown in Algorithm~\ref{alg:max_heap}. Each function's execution time at the minimum memory configuration i.e., $128$MB is calculated and is used for constructing the max-heap. We store the execution time of the function at a particular configuration as the node value and the function name and its memory configuration are further saved as the node's metadata (Line 5-8). The function at a particular memory configuration having the highest execution time will be automatically stored at the head of the max-heap tree (Line 9). We first check if this base configuration satisfies the SLO requirements. In case it does, we stop the iteration and return the configuration (Line 11-13). Otherwise, in the next step, we pop the head from the max-heap (Line 14), increase its memory to decrease its execution time (Line 16) and then push the function again back to the heap with the updated memory and execution time (Line 17-20). After this update, we check if the configuration satisfies the SLO requirements. In case it does, we stop the iteration and return the configuration (Line 11-13). Otherwise, we continue the process by popping the function at the head until a configuration is found. If no configuration is found, an empty dictionary is returned. The overall complexity of this approach is given by:

\begin{equation}
O(NM{\log N})\label{eq:heapcom} 
\end{equation}

This method is highly scalable and also does locally optimal steps to lower the overall execution time of function call. 

\begin{algorithm}[t]
\small
\SetAlgoLined
\KwInput{ func\_list, mem\_config\_list: List[ ], SLO)
                  
}
  \KwOutput{result\_config = Dict[str, int]}
  
min\_mem\_config = \textbf{min}(mem\_config\_list)

\For{func\_name in func\_list}{
    \tcp{init minimum memory assignment for all functions}
    res\_config$\left[\text{func\_name}\right]$ = min\_mem\_config\;
}

\tcp{prepare heap with function's exec time at min memory }
\For{fname in func\_list}{

    func\_exec\_time = \textbf{exec\_time}(fname, min\_mem\_config)\;
    func\_heap.\textbf{append}({func\_exec\_time, fname})\;
    }

\textbf{heapify\_max}(func\_heap)\tcp*[l]{reorder heap}

\Do{func\_heap is not empty}{ 
    
    \tcp{check for objective(s) satisfaction.}
    \If{  \textbf{estimate\_exec\_time}(res\_config) $\leq$ SLO}{
        return res\_config\;
    }
    top\_func = \textbf{heappop\_max}(func\_heap)\;
    \If{not \textbf{all\_memory\_config\_evaluated}(top\_func)}{
            \tcp{update memory and time, then append to heap}
            func\_new\_mem = \textbf{update\_memory}(top\_func)\;
            func\_new\_exec\_time = \textbf{get\_exec\_time}(top\_func, func\_new\_mem)\;
            func\_heap.append({func\_new\_exec\_time, top\_func})\;
                
            res\_config[top\_func] = func\_new\_mem\tcp*[l]{update}
            \textbf{heapify\_max}(func\_heap)\tcp*[l]{reorder heap}

    }
    
}   
return   \tcp*[l]{return the empty config}  
    
\caption{SLAM-SLO Algorithm}\label{alg:max_heap}
\end{algorithm}

\textbf{SLAM-SLO-Min-Cost}:
We further modified the \textit{SLAM-SLO} algorithm to take cost into account for finding the optimal configuration with the \ac{moc} as the objective along with the SLO requirements. Here, the algorithm uses the \textit{SLAM-SLO} found optimal configuration as the default configuration and tries to optimize on top of it for finding minimum cost configuration. In this, every time we pop the function from the head of max-heap, we check for the following inequality at the new updated memory for that function:
\begin{equation}
\abs{\frac{\text{new\_cost} - \text{old\_cost}}{\text{old\_cost}}} \le \abs{\frac{\text{old\_exec\_t}-\text{new\_exec\_t}}{\text{old\_exec\_t}}}\label{eq:optimal} 
\end{equation}

where new\_cost and new\_exec\_t are the cost and execution time of an application invocation after updating the memory of the function, and old\_cost and old\_exec\_t correspond to the cost and execution time before the update. In case the inequality holds, we put the function back into the max-heap with the updated execution time. In case it doesn't, we fix the memory for that function in the final configuration. This also allows us to reduce the search space for finding the configuration satisfying the minimum cost objective.

\textbf{SLAM-SLO-Min-Time}: This modified version of the \textit{SLAM-SLO}  algorithm also uses the \textit{SLAM-SLO} found optimal configuration as the default configuration and tries to optimize on top of it for finding minimum execution time configuration. It then leverages the binary search algorithm to find the configuration with minimum time. It uses the \textit{SLAM-SLO} found optimal configuration execution time ($\beta$ in seconds) as the maximum time and $0s$ as the minimum time. It then updates the SLO requirement to the middle of maximum and minimum time and calls the \textit{SLAM-SLO} algorithm to find an optimal configuration. If a configuration is found, then the maximum is set to the execution time for that configuration, otherwise the minimum is updated to the previously found middle. This way it continues until a configuration is found with minimum application execution time. In order not to run the binary search indefinitely, we use a hyperparameter called precision ($\gamma$). When the lower and upper execution time bounds get closer than the precision hyperparameter, we stop the search and return the attained configuration. As a default value of the parameter, we chose $\gamma=0.01s$, which can be easily changed. The complexity of the algorithm is given by: 

\begin{equation}
O\left(NM{\log N}  \times \log \left(\frac{\beta}{\gamma}\right)\right)\label{eq:SLAM-SLO-Min-Time} 
\end{equation}


\begin{figure*}[t]
\centering
\begin{subfigure}{0.17\textwidth}
   \centering
    \includegraphics[width=0.9\linewidth]{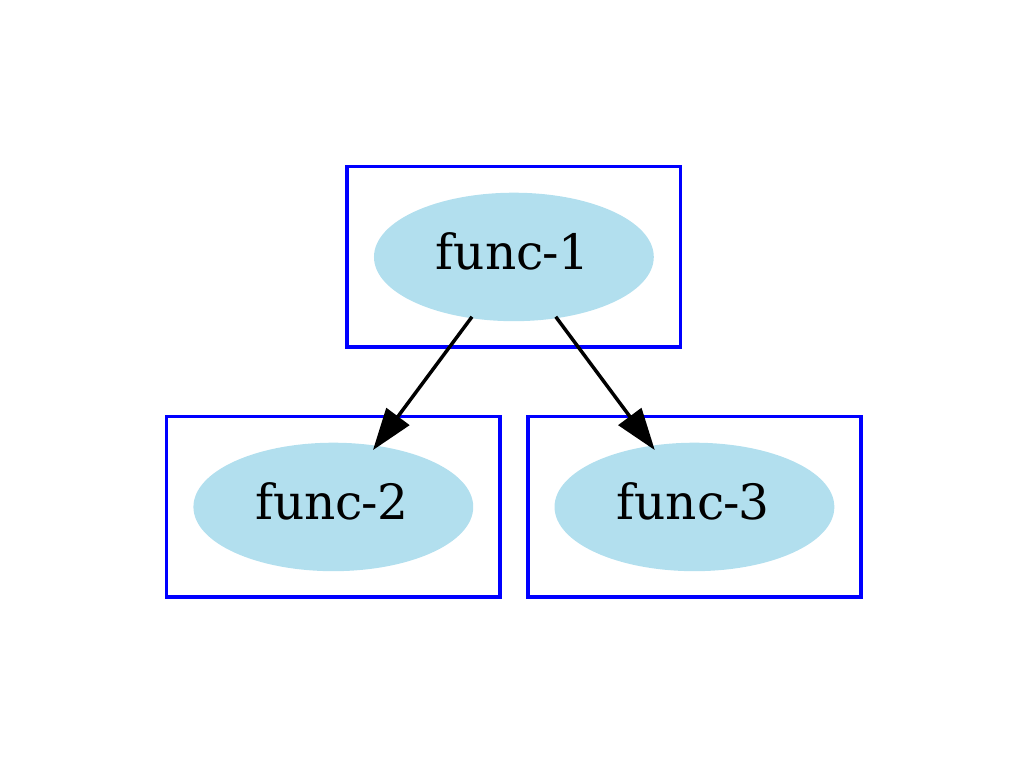}
    \caption{3-functions test app} \label{fig:3_synthetic}
\end{subfigure}\hfill
\begin{subfigure}{0.17\textwidth}
    \centering
    \includegraphics[width=0.91\linewidth]{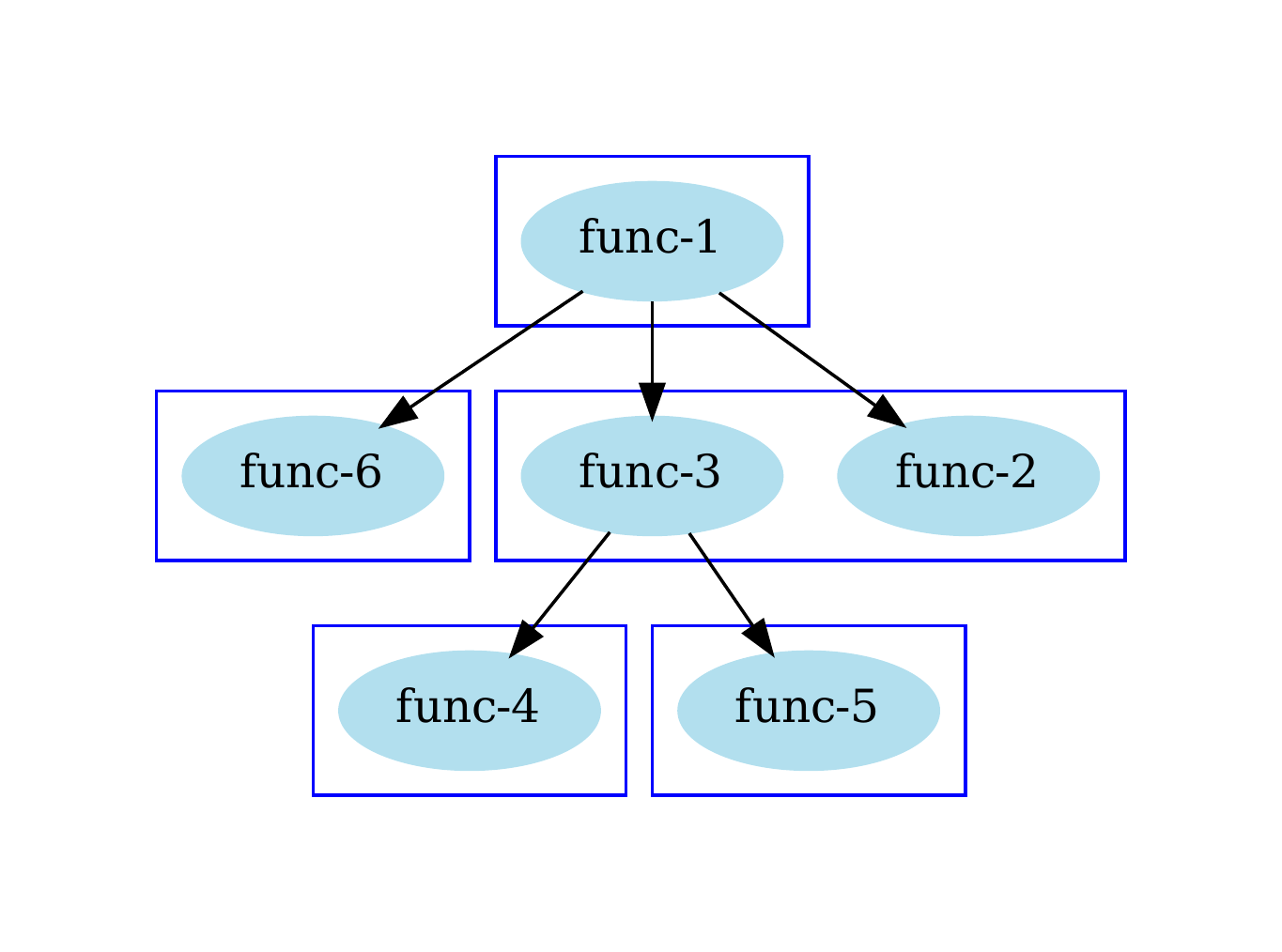}
    \caption{6-functions test app} \label{fig:6_synthetic}
\end{subfigure}\hfill
\begin{subfigure}{0.29\textwidth}
   \centering
    \includegraphics[width=1\linewidth]{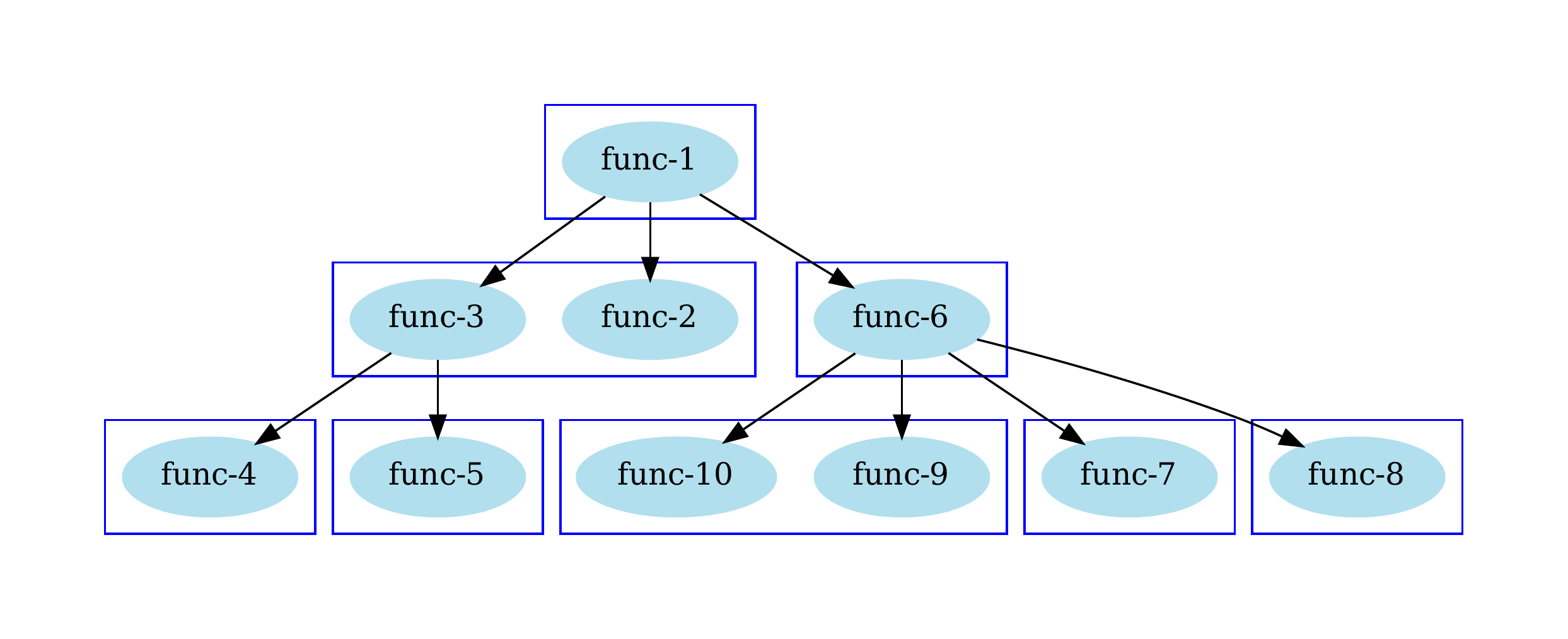}
    \caption{10-functions test app} \label{fig:10_synthetic} 
\end{subfigure}\hfill
\begin{subfigure}{0.34\textwidth}
   \centering
    \includegraphics[width=1\linewidth]{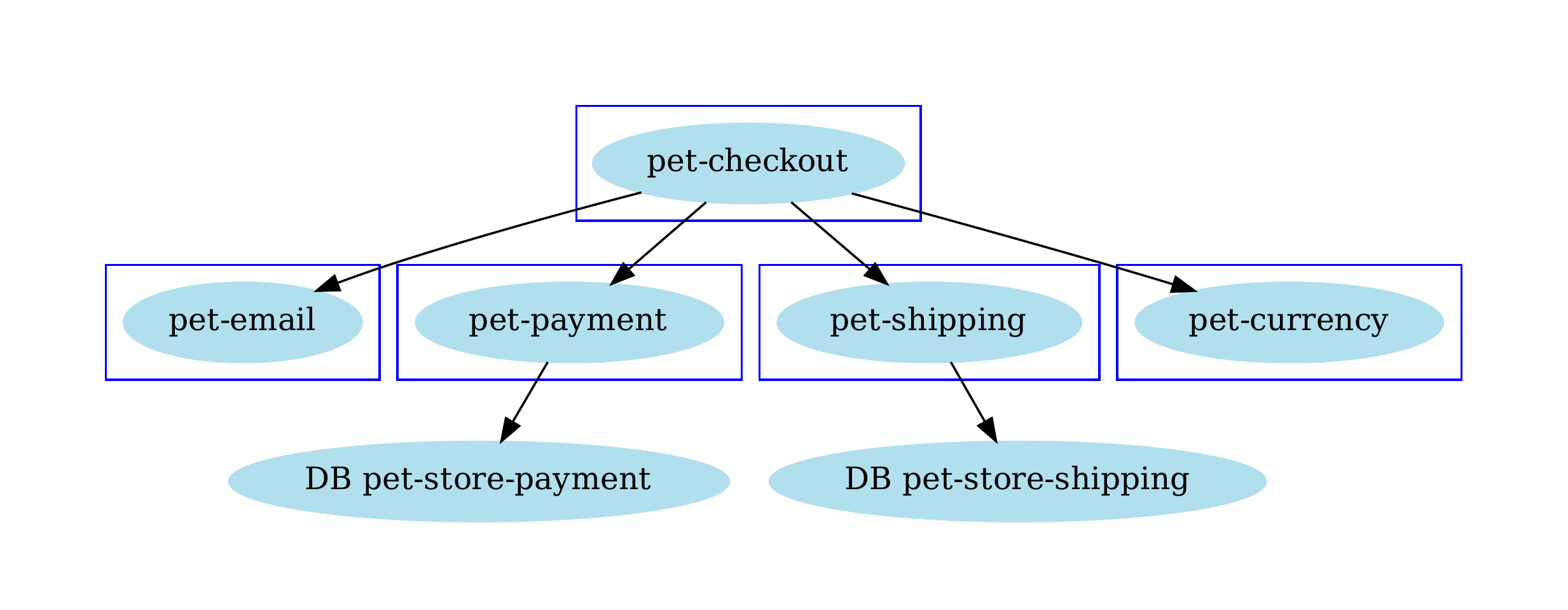}
    \caption{Real-world based app} \label{fig:RealCallGraphpdf} 
\end{subfigure}

\caption{
Call graphs for the applications used for evaluating \textit{SLAM}.}
\label{fig:dependency_call_graphs}
\end{figure*}
\section{Evaluation Settings}
\label{sec:evaluation_settings}
We test the proposed \textit{SLAM} tool for serverless applications deployed on AWS Lambda, a popular serverless cloud platform. \textit{SLAM} tool itself was run on a machine with $8$ physical cores (Intel(R) Core(TM) i7-8550U CPU @ 1.80GHz CPU) with hyperthreading enabled  and $16$ GB of RAM. These conditions are similar to a typical cloud VM.

SLAM has a default list of memory configuration values ($mem\_config\_list$ in Table~\ref{tab1:symbols}), that it chooses from when generating memory configurations for the application. As all the functions within our test applications use only one thread, we limit the maximum memory configuration to $2$GB, since as at that point AWS stops increasing the portion of the allocated vCPU and increases the number of available vCPU~\cite{wang2018peeking}, which will then not used by the application. For our experiments, total number of requests for load generation is set to as \texttt{50}. 

\subsection{Test Applications}
\label{sec:test_applications}
\subsubsection{Synthetic Applications}
To test the \textit{SLAM} tool, we have developed an interface that can create automatically synthetic applications having a different number of functions. The input to the interface defines the application call tree containing functions that are either invoked in parallel or sequence. This way we can generate complex applications with as many functions. Such a structure gives us the opportunity to test the limits of the \textit{SLAM} tool and understand how much error is accumulated if the application contains many cloud functions with a combination of sequence and parallel invocations. 
Each function within the application is a compute-intensive function which
calculates the sum of remainders for $N$ when divided by all numbers between $2$ and $N$, where $N$ is the parameter fixed for the function. The simplicity of the algorithm allows us to simulate test applications with heterogeneous functions requiring different compute/memory resources by scaling $N$. Each function within the application has a different value for $N$ and is assigned randomly. An example application with three functions where one function (func-1) is invoking the other two (func-2, func-3) in the sequence is created, and its call graph is shown in Figure~\ref{fig:3_synthetic}.



We additionally created two more synthetic complex applications containing $6$ and $10$ functions incorporating sequence and parallel invocations to test the \textit{SLAM} tool. Their call graphs are shown in Figure~\ref{fig:6_synthetic} and Figure~\ref{fig:10_synthetic} respectively. Functions in the same box are called in parallel to each other, while the ones on the same level are called in sequence. The directed edges show the function which has generated the invocation for the other functions on the lower level.  

Such complex application call graphs allow us to estimate how well \textit{SLAM} adapts to changing execution time when a change in a leaf function's configuration affects the execution time of the other higher level functions. 


 
 

\subsubsection{Real-world based Application}
Since the synthetic application workloads do not fully represent the real-world use cases for serverless applications, therefore we created a pet store application based on an open-source spring-based application\footnote{https://github.com/spring-projects/spring-petclinic} consisting of five FaaS functions and two NoSQL databases. Its call graph is shown in Figure~\ref{fig:RealCallGraphpdf}. We used DynamoDB for the two NoSQL databases. This application is special, since the functions querying databases will not have any influence on execution time with the increase in memory. 

In this application, when the client selects a pet in the front-end for buying, it first automatically invokes the \textit{pet-checkout} function, which in turn is responsible for getting all the details needed for the purchase by invoking other functions. First, it calls the \textit{pet-currency} function to convert the pet price to USD. Then it calls the \textit{pet-payment} service for the client to pay for the pet. If the payment is successful then the \textit{pet-checkout} function will invoke \textit{pet-shipping} which will log the pet shipping details in the database. After successful completion of all the previous steps, the final \textit{pet-email} function is called, which generates a summary email and sends it to the client. The application is just a skeletal representation of what a real one would look like; it does not ship anything.


\section{Evaluation}
\label{sec:evaluation}
We design our experiments to answer the questions:
\begin{itemize}
    \item \textbf{Q1. \textit{SLAM} estimation time accuracy}: how accurate is \textit{SLAM} in estimating the execution time of an application for the given or found configuration at different SLOs?
    
    \item \textbf{Q2. \textit{SLAM} configuration finding accuracy}: how accurate is \textit{SLAM} in finding the configuration satisfying the given SLOs and objectives for an application?
  
  \item \textbf{Q3. \textit{SLAM} configuration finding efficiency and scalability}: how efficient is \textit{SLAM} in finding the configuration satisfying the given SLOs and objectives for an application? Additionally, how does the \textit{SLAM} tool scale with the increase in the number of functions of the application?
  
\end{itemize}

\subsection{Q1. SLAM estimation time accuracy}
\label{sec:est_time_accuracy}
\begin{figure*}[t]
\centering
\begin{subfigure}{0.24\textwidth}
   \centering
    \includegraphics[width=1\linewidth]{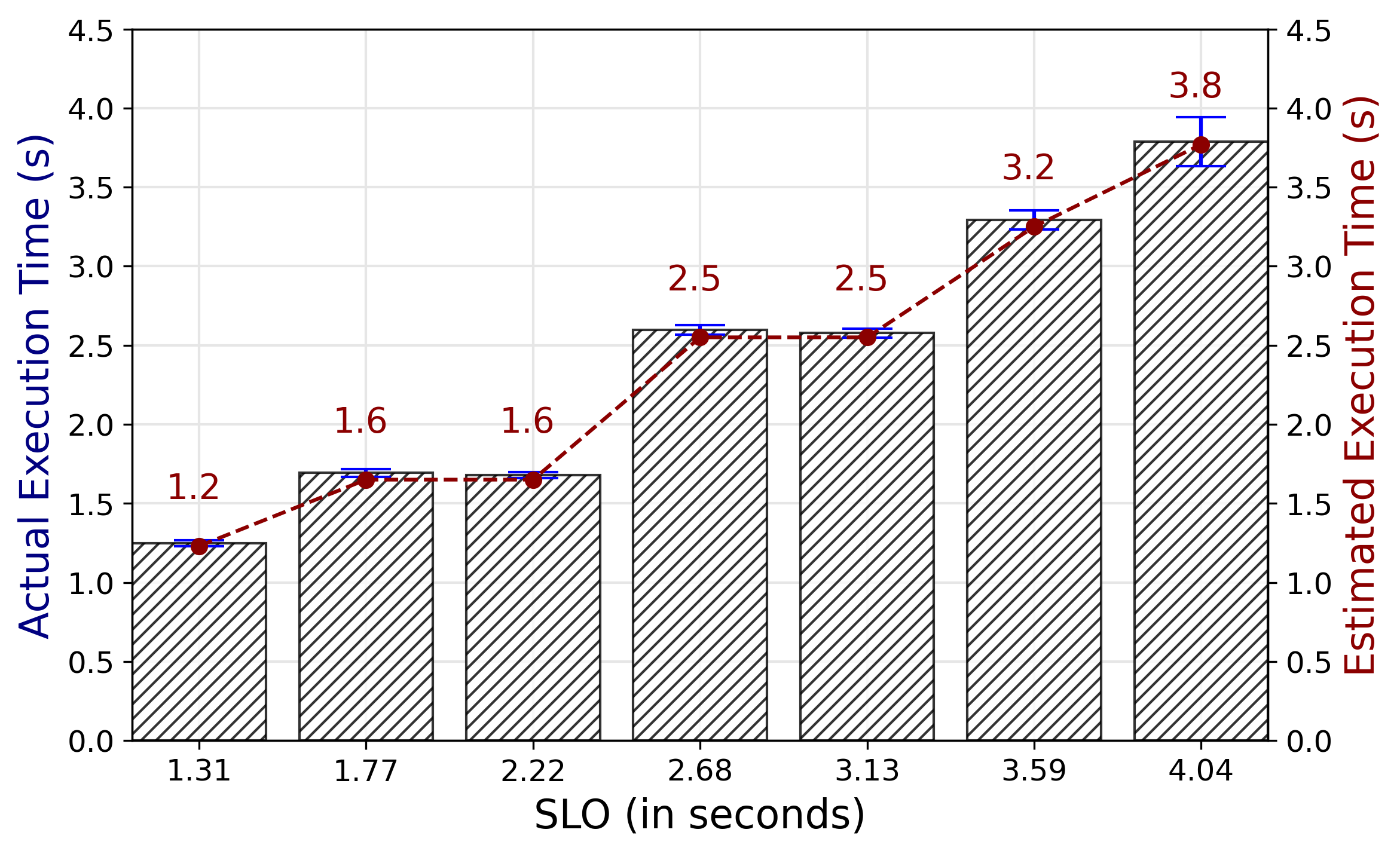}
    \caption{3-functions test application} \label{fig:3SLO_Estim}
\end{subfigure}\hfill
\begin{subfigure}{0.24\textwidth}
    \centering
    \includegraphics[width=1\linewidth]{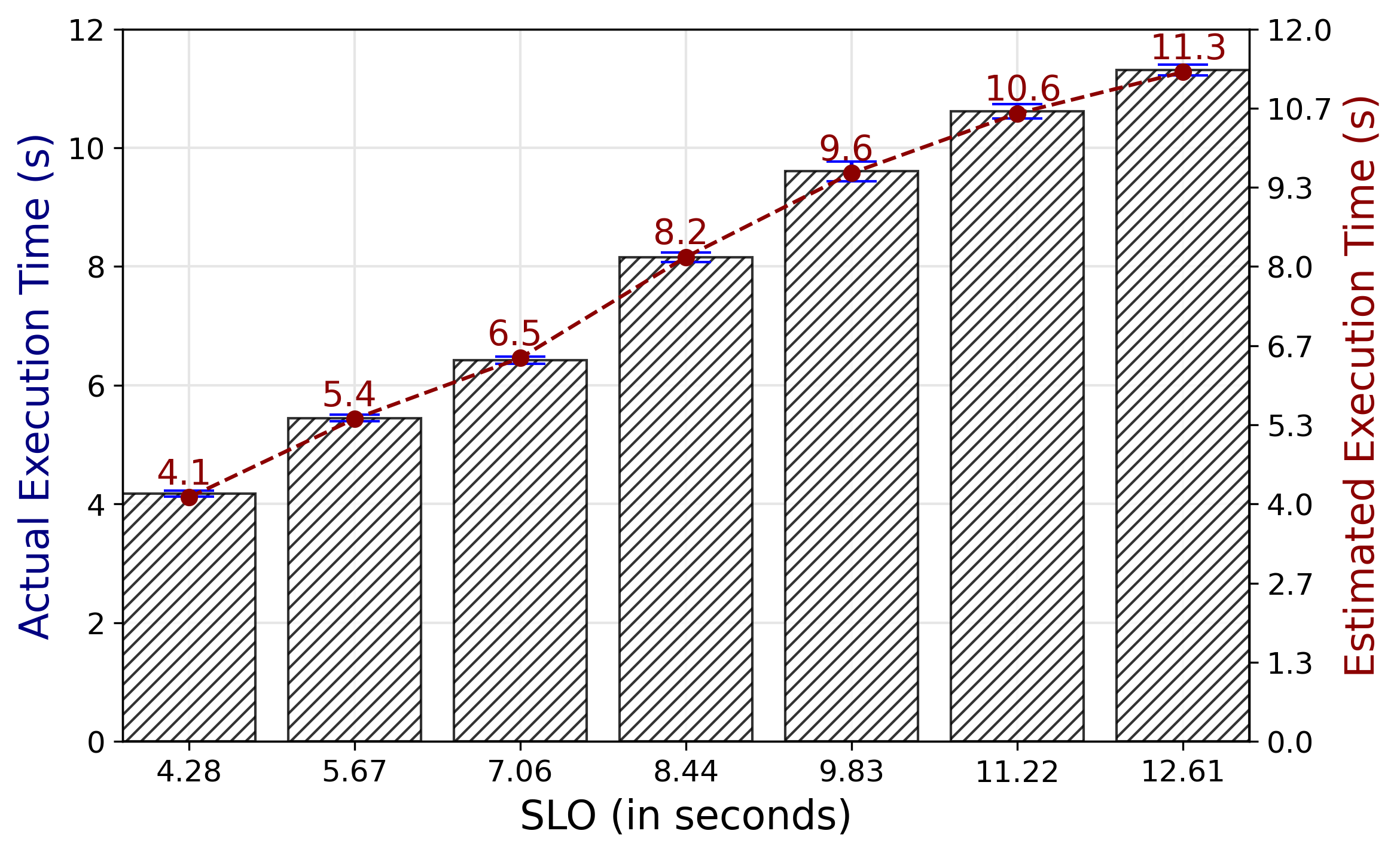}
    \caption{6-functions test application} \label{fig:6SLO_Estim}
\end{subfigure}\hfill
\begin{subfigure}{0.24\textwidth}
   \centering
    \includegraphics[width=1\linewidth]{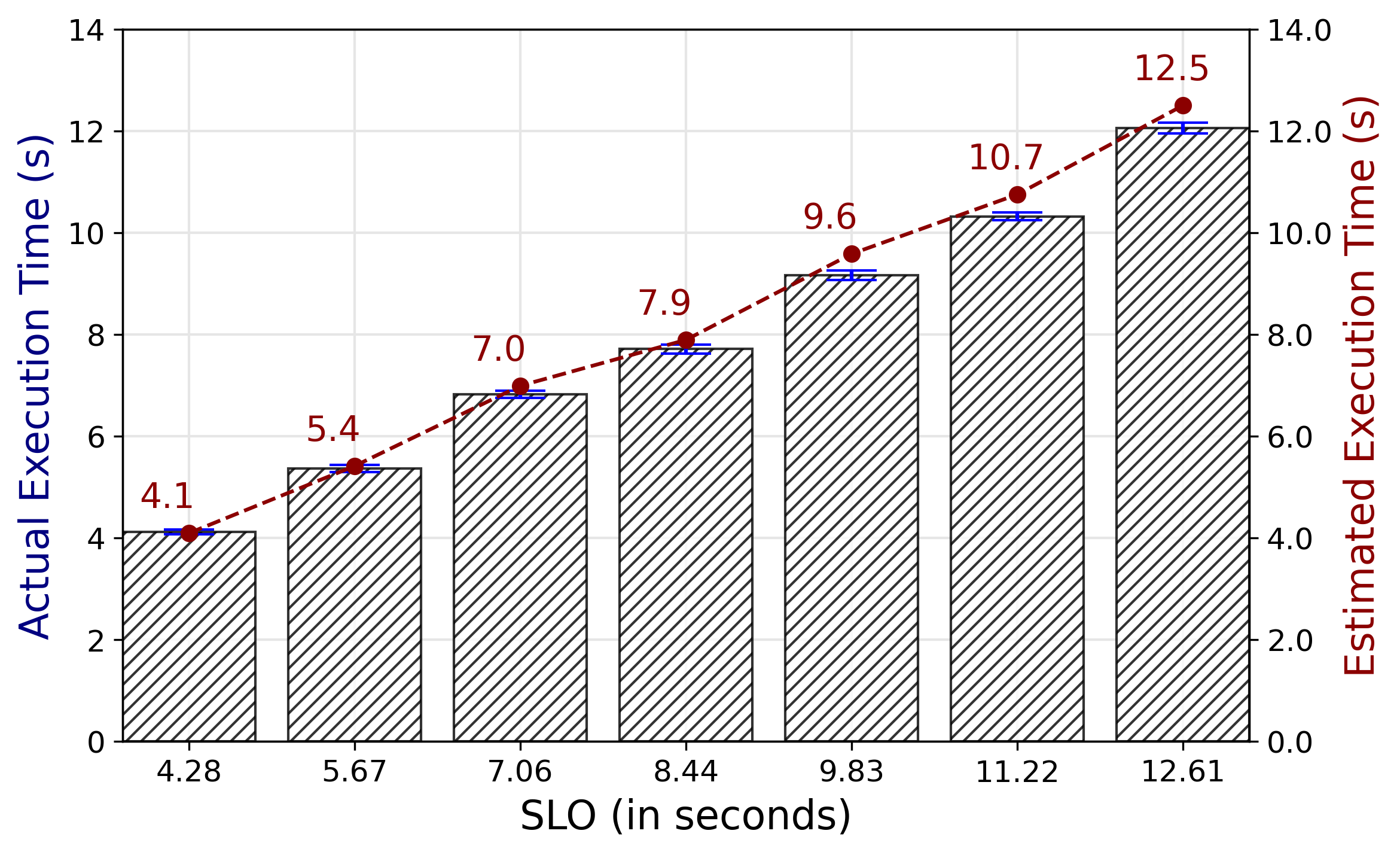}
    \caption{10-functions test application} \label{fig:10SLO_Estim} 
\end{subfigure}
\begin{subfigure}{0.24\textwidth}
   \centering
    \includegraphics[width=1\linewidth]{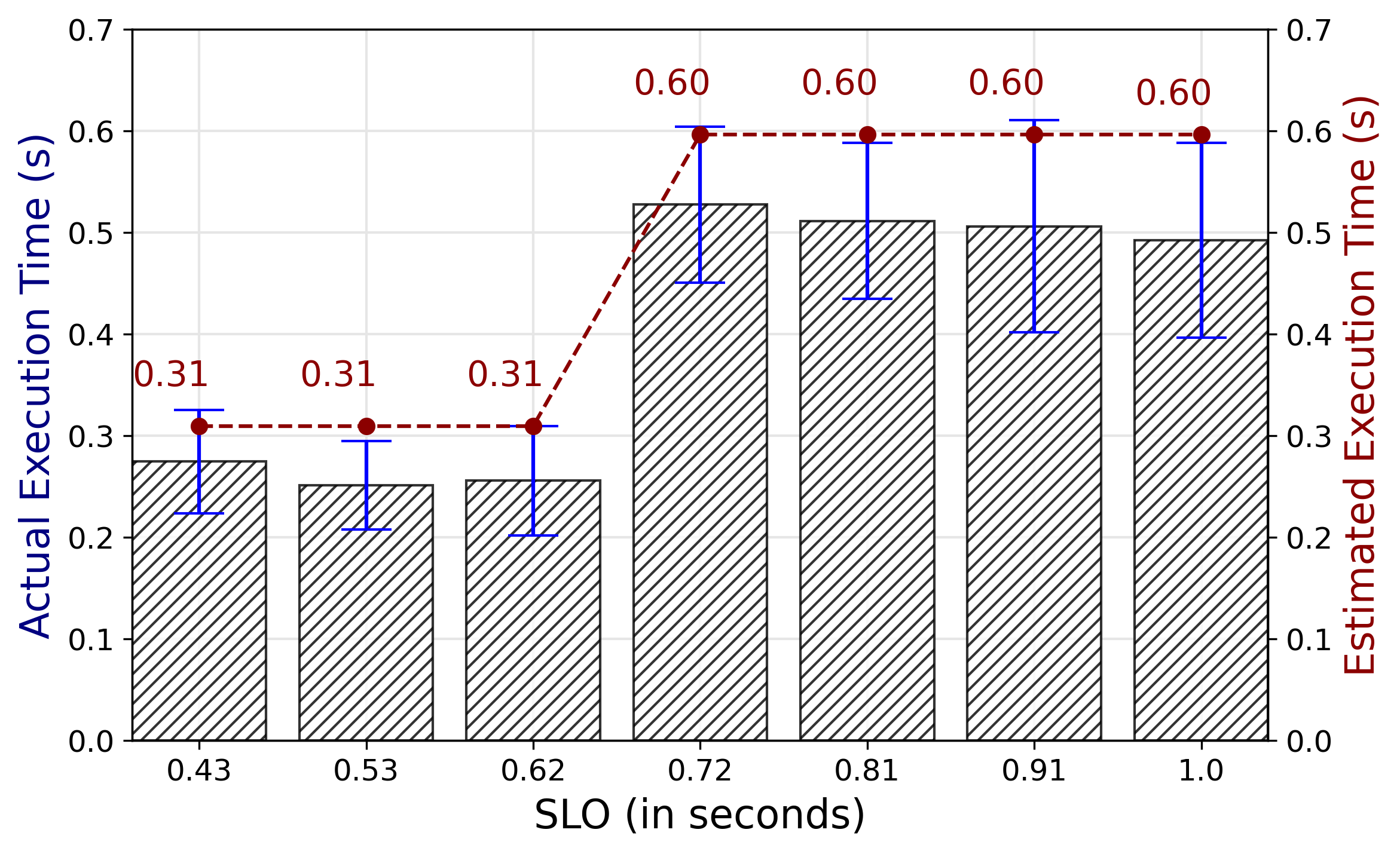}
    \caption{Real-world based application} \label{fig:RealSLO_Estim} 
\end{subfigure}
\caption{
Actual execution time box plot overlaid with the estimated execution time by \textit{SLAM} run with different SLOs. }
\label{fig:estimation_actual_dist}
\end{figure*}

\begin{figure*}[t]
\centering
\begin{subfigure}{0.24\textwidth}
   \centering
    \includegraphics[width=1\linewidth]{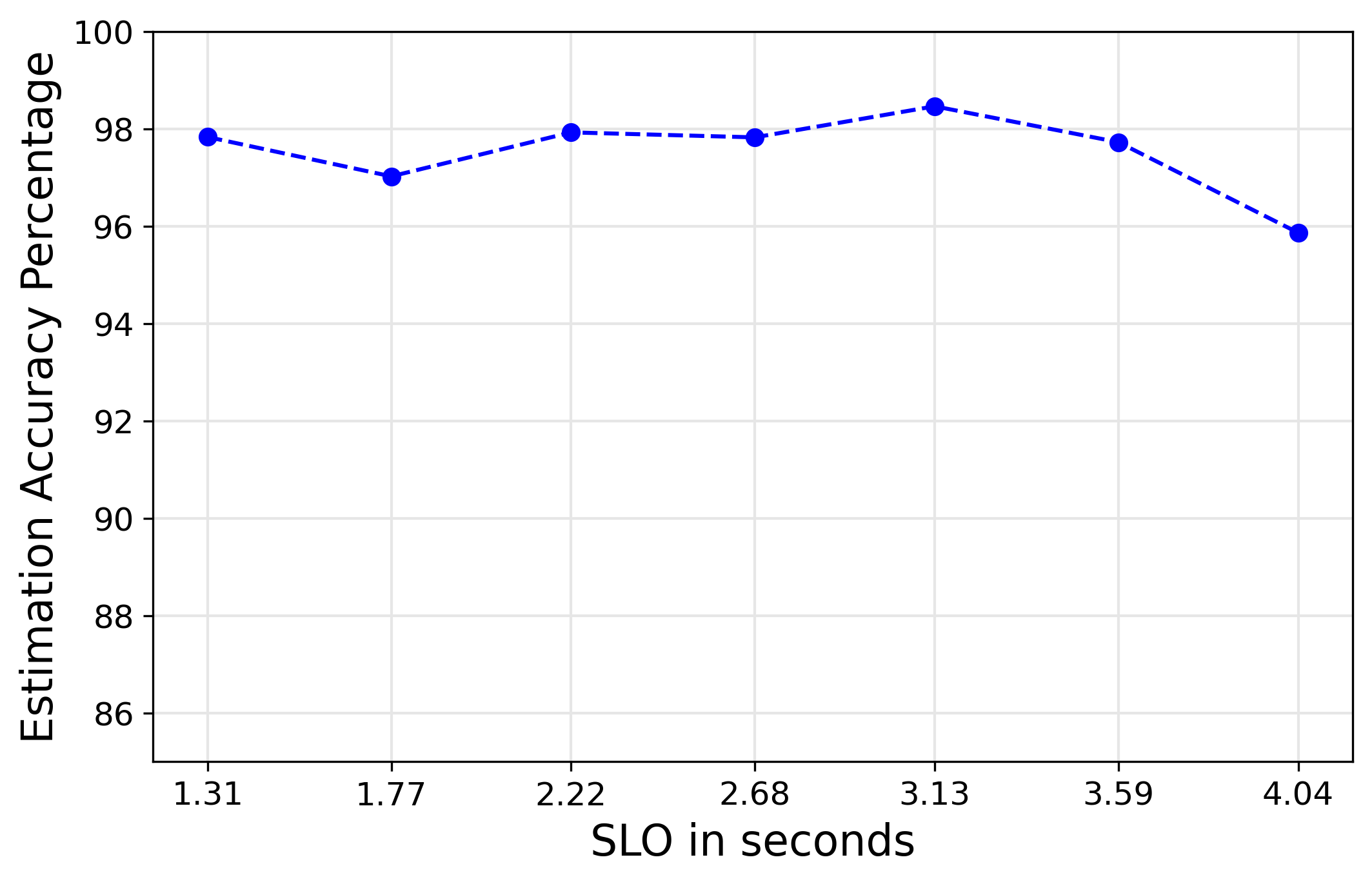}
    \caption{3-functions test application} \label{fig:3Error}
\end{subfigure}\hfill
\begin{subfigure}{0.24\textwidth}
    \centering
    \includegraphics[width=1\linewidth]{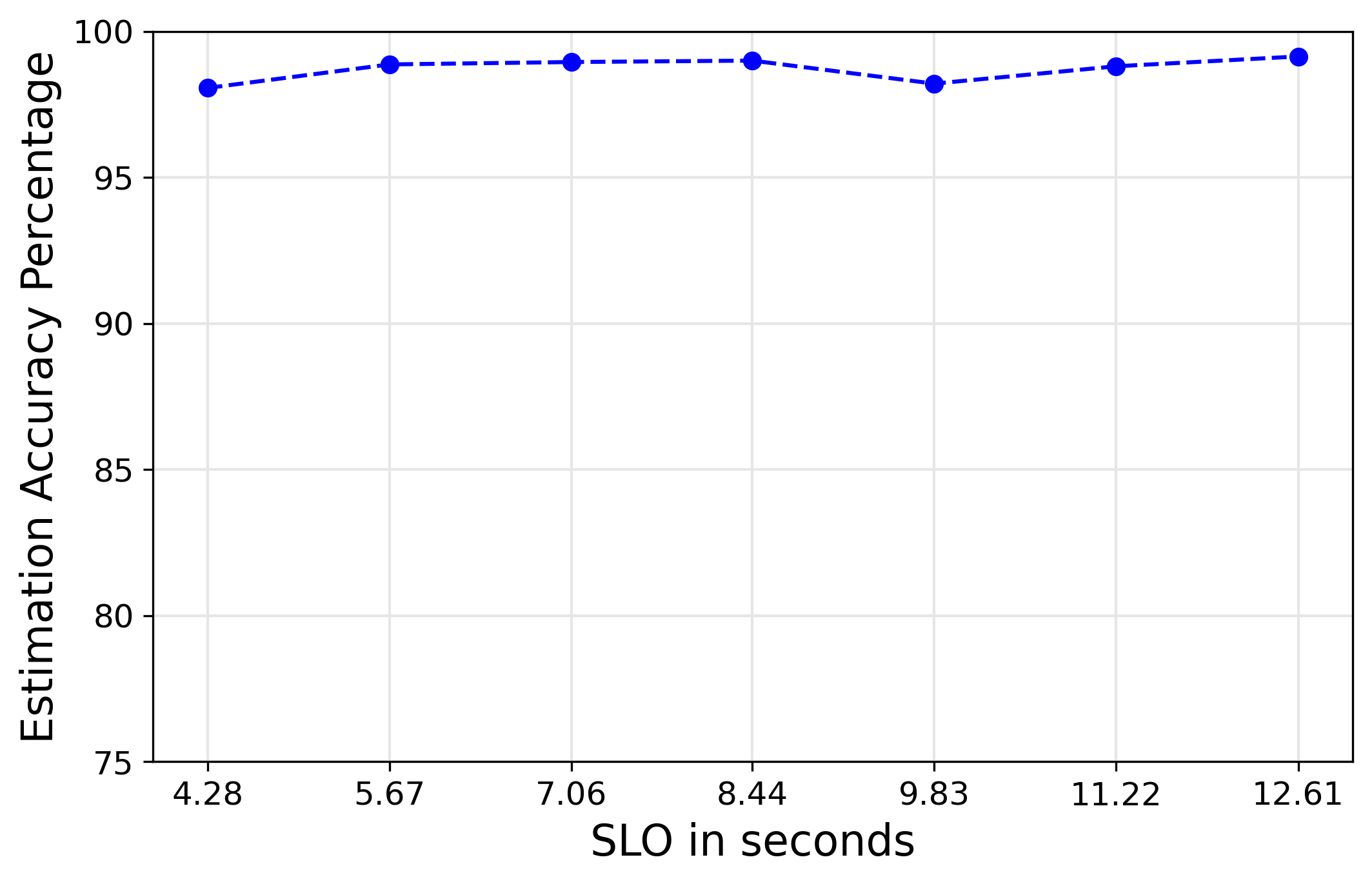}
    \caption{6-functions test application} \label{fig:6Error}
\end{subfigure}\hfill
\begin{subfigure}{0.24\textwidth}
   \centering
    \includegraphics[width=1\linewidth]{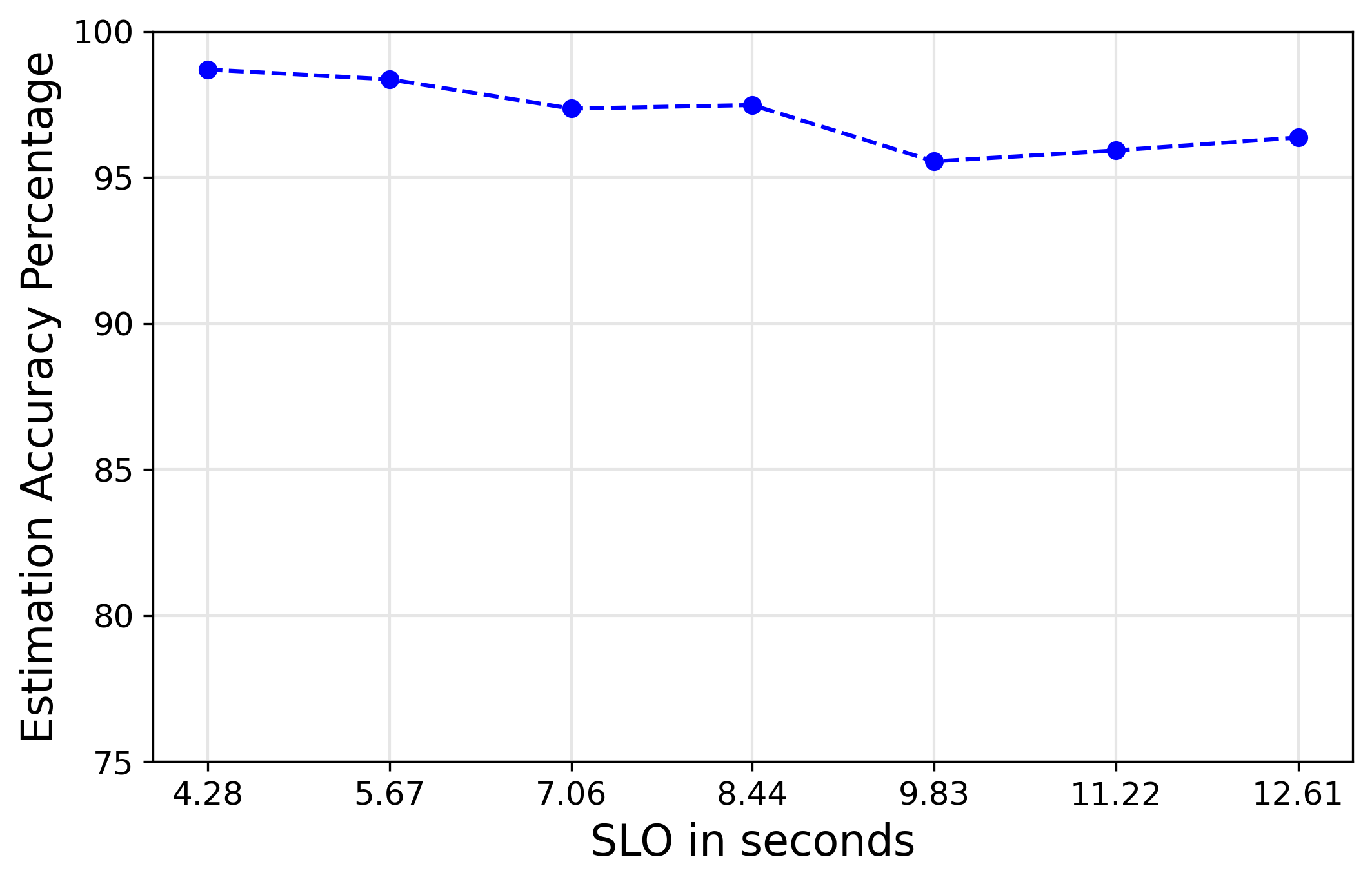}
    \caption{10-functions test application} \label{fig:10Error} 
\end{subfigure}
\begin{subfigure}{0.24\textwidth}
   \centering
    \includegraphics[width=1\linewidth]{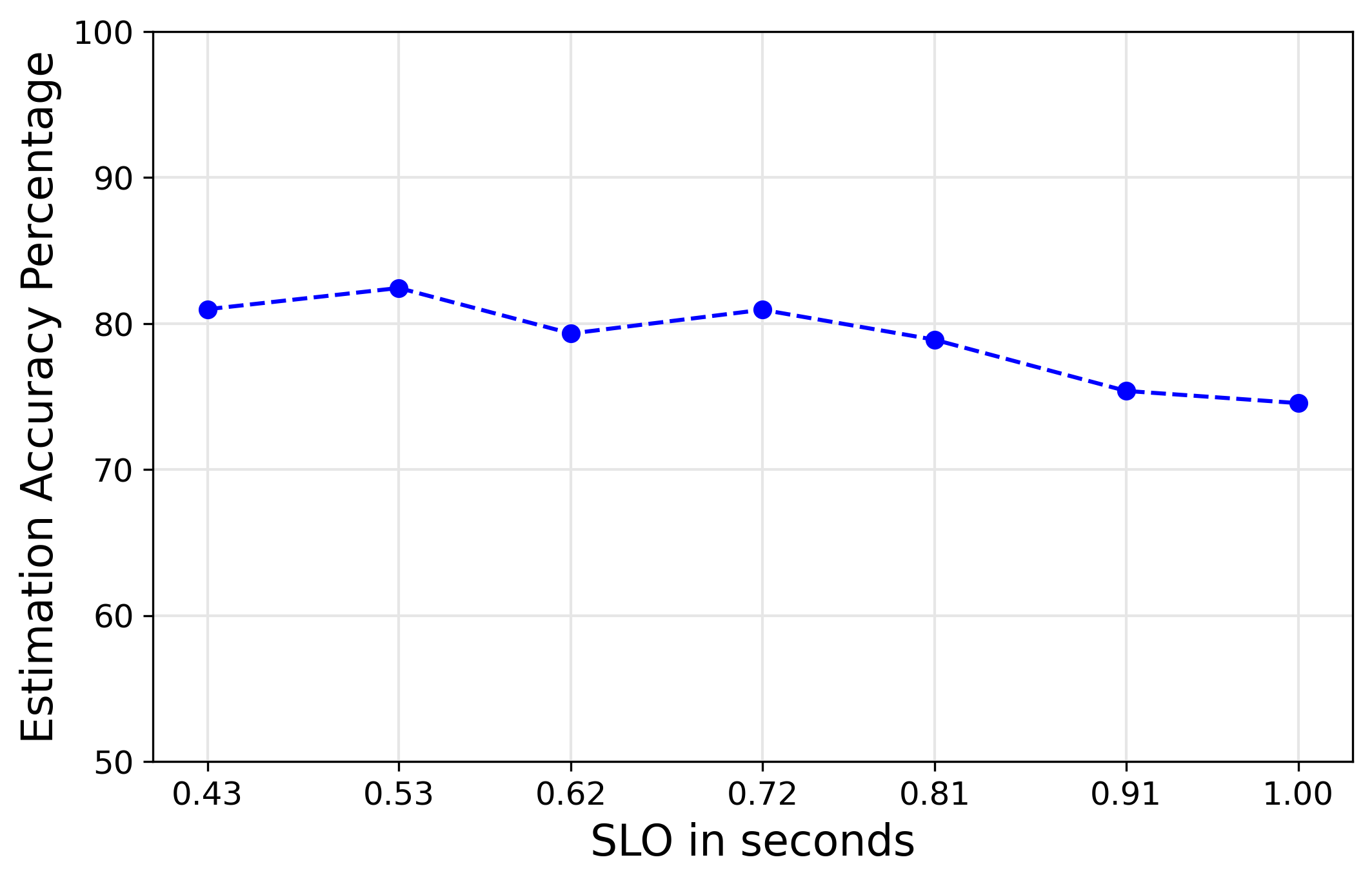}
    \caption{Real-world based application} \label{fig:RealError} 
\end{subfigure}
\caption{
Execution time estimation accuracy percentage for the four test applications at different SLOs.}
\label{fig:mspe_error_applications}
\end{figure*}


To demonstrate the effectiveness of the \textit{SLAM} tool in estimating the execution time of the application, we test it on three synthetic and one real-world-based application. For this test, \textit{SLAM} tool's \textit{SLAM-SLO} algorithm is used to find the memory configurations for the given different SLOs without any additional objectives. Based on the found configuration, we then configured all the functions with the memory values suggested by \textit{SLAM-SLO} and invoke the serverless application $100$ times to get the actual application's execution time distribution.
Figure~\ref{fig:estimation_actual_dist} shows the actual experiment execution time box plot overlaid with the estimated execution time by \textit{SLAM-SLO} algorithm for all four test applications at different SLOs when configured with the found memory configurations. 

Additionally,  we measured the execution time estimation accuracy percentage for the four test applications at different SLOs and is shown in Figure~\ref{fig:mspe_error_applications}. For computing the accuracy at different SLOs, we calculate the mean squared percentage error between the estimated and actual execution time for the found configuration and then subtract it from $100$. 

Next, we discuss the results of the two classes of the test applications in more detail. 

\subsubsection{Synthetic Applications}
From the Figure~\ref{fig:estimation_actual_dist}, one can observe that in the three synthetic applications the estimated execution time is either lower or equal to that of the specified SLOs. Additionally, from the overlaid graph of estimated execution time in Figure~\ref{fig:estimation_actual_dist}, we can observe that the estimated execution time to a great extent is closer to the actual execution time at different SLOs. To verify it further, in Figure~\ref{fig:mspe_error_applications}, the measured execution time estimation accuracy percentage for the three test applications at different SLOs is above 90\%. 
\subsubsection{Real-world based Application}
From the overlaid graph of estimated execution time in Figure~\ref{fig:RealSLO_Estim}, one can observe that the estimated execution time is a bit higher than the actual execution time at different SLOs which is also evident from the Figure~\ref{fig:RealError} where the measured execution time estimation accuracy percentage at different SLOs is lower as compared to synthetic applications (ranging between 70\% and 85\%), but similar to the three synthetic applications, the estimated execution time for this application is also either lower or equal to that of the specified SLOs. Thus, the configuration selected by the \textit{SLAM} tool is good enough to fulfill the desired SLOs. 

One reason for the higher estimated execution time at different SLOs could be due to the high variance in the actual execution time of the functions within the application (as seen in Figure~\ref{fig:dist128Baas}) because of the involvement of components such as DynamoDB which can lead to the variable execution time of the application. Moreover, the overall execution time of this application is smaller as compared to synthetic applications and thus even the small inherent variance within the application can cause high relative error rates and hence the drop in the estimation of the accuracy.  Nonetheless as mentioned earlier, the configuration selected by the \textit{SLAM} tool is good enough to fulfill the desired SLOs.


\subsection{Q2. \textit{SLAM} configuration finding accuracy}
\label{sec:config_finding_accuracy}
In this experiment, for determining the accuracy of \textit{SLAM} in finding the configuration at the given SLOs, we have considered two aspects presented next. 
\subsubsection{Precision of requests conforming SLO requirements}

\begin{figure}[t]
\centering
\begin{subfigure}{0.49\columnwidth}
   \centering
    \includegraphics[width=1\linewidth]{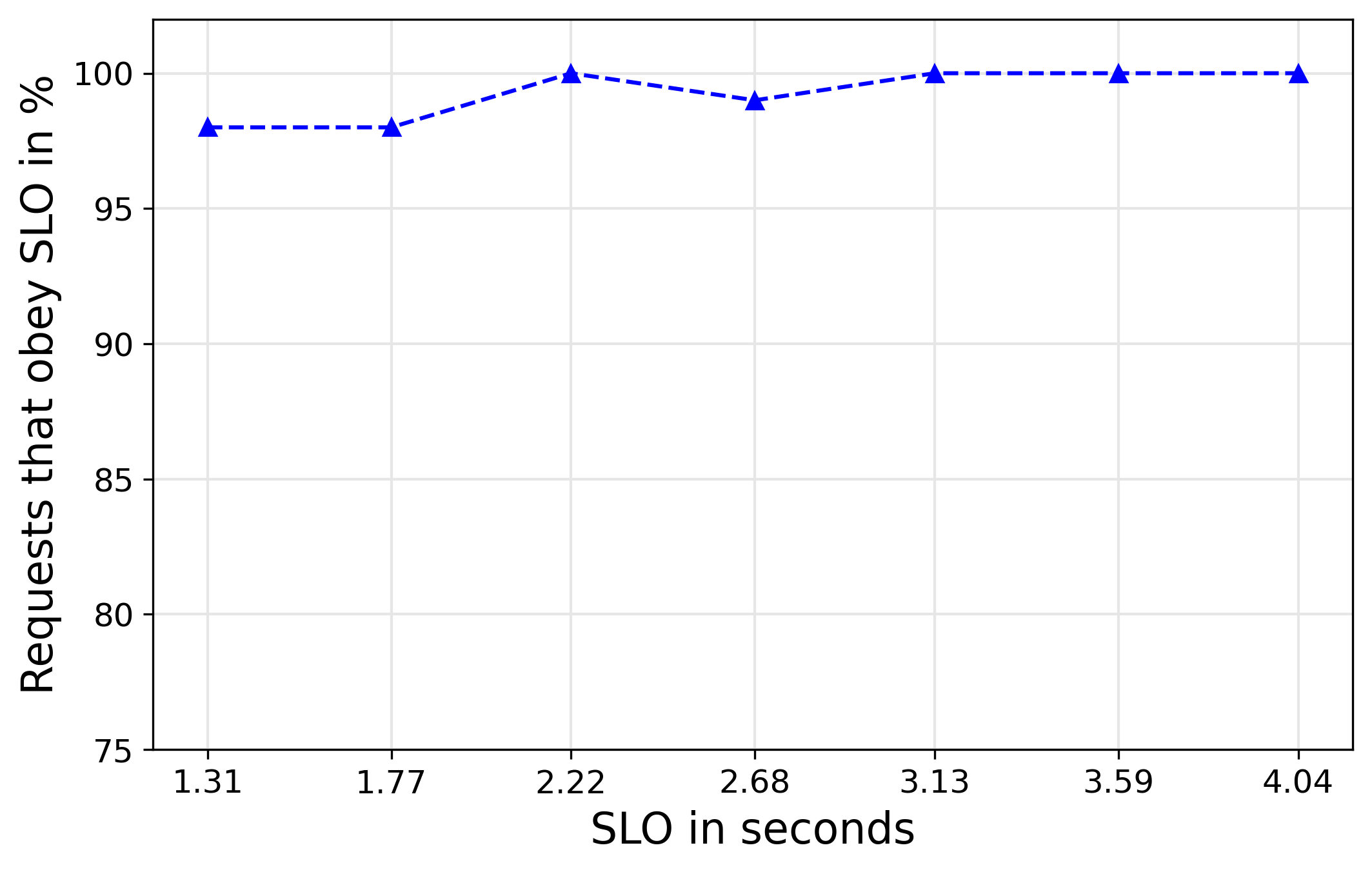}
    \caption{3-functions test application} \label{fig:2Prercision}
\end{subfigure}\hfill
\begin{subfigure}{0.49\columnwidth}
    \centering
    \includegraphics[width=1\linewidth]{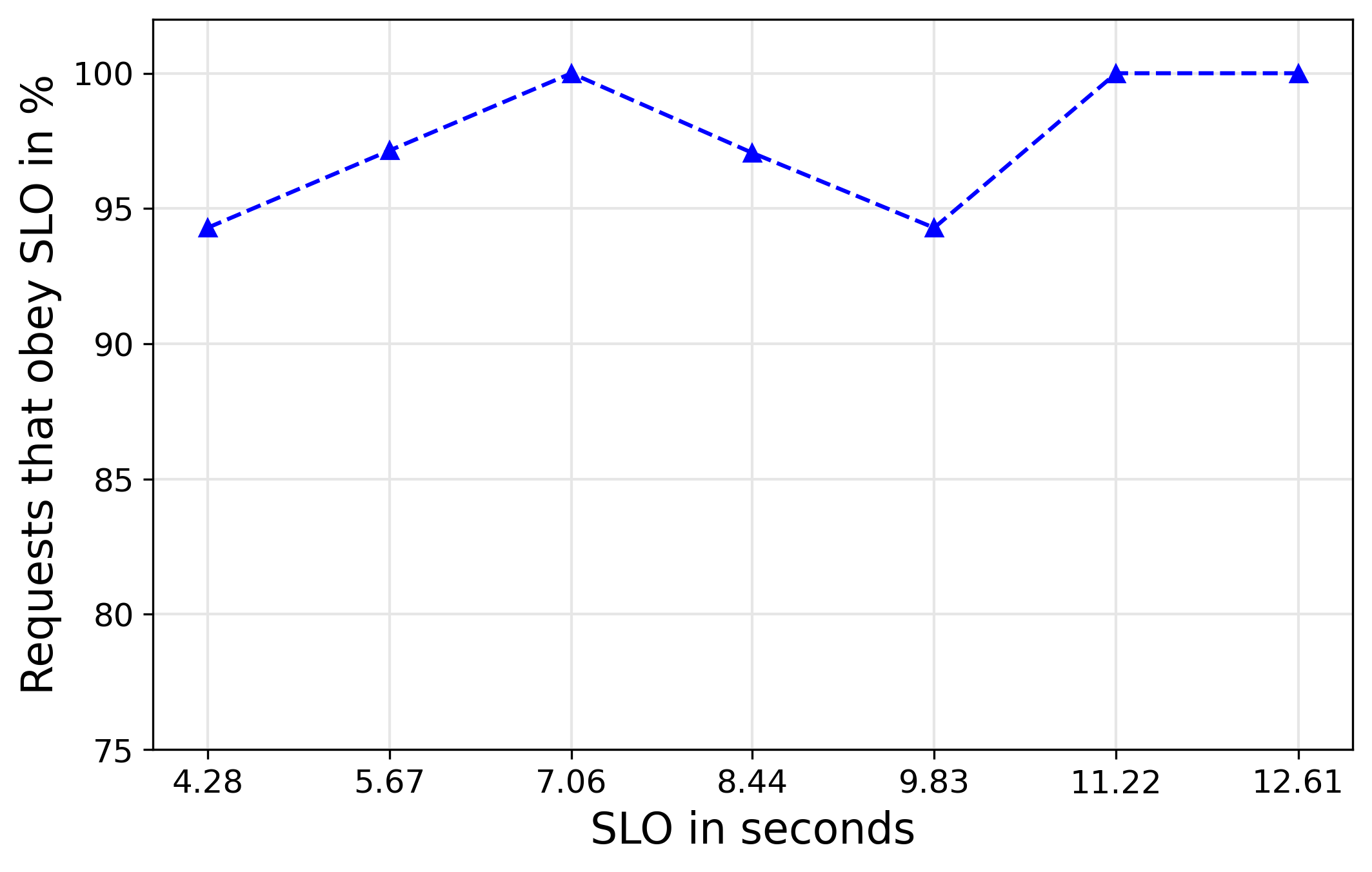}
    \caption{6-functions test application} \label{fig:6Prercision}
\end{subfigure}\hfill
\begin{subfigure}{0.49\columnwidth}
   \centering
    \includegraphics[width=1\linewidth]{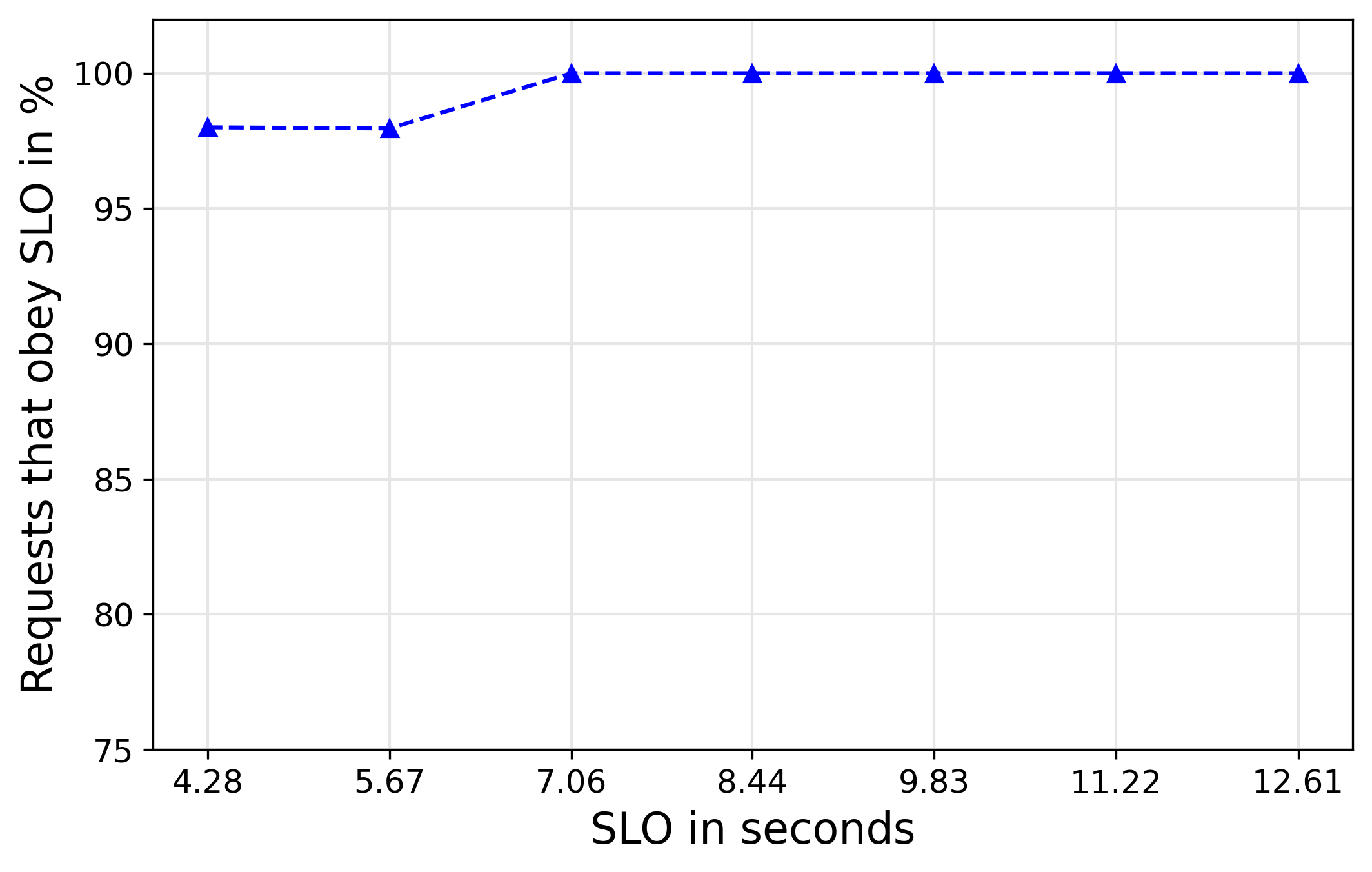}
    \caption{10-functions test application} \label{fig:10Prercision} 
\end{subfigure}\hfill
\begin{subfigure}{0.49\columnwidth}
   \centering
    \includegraphics[width=1\linewidth]{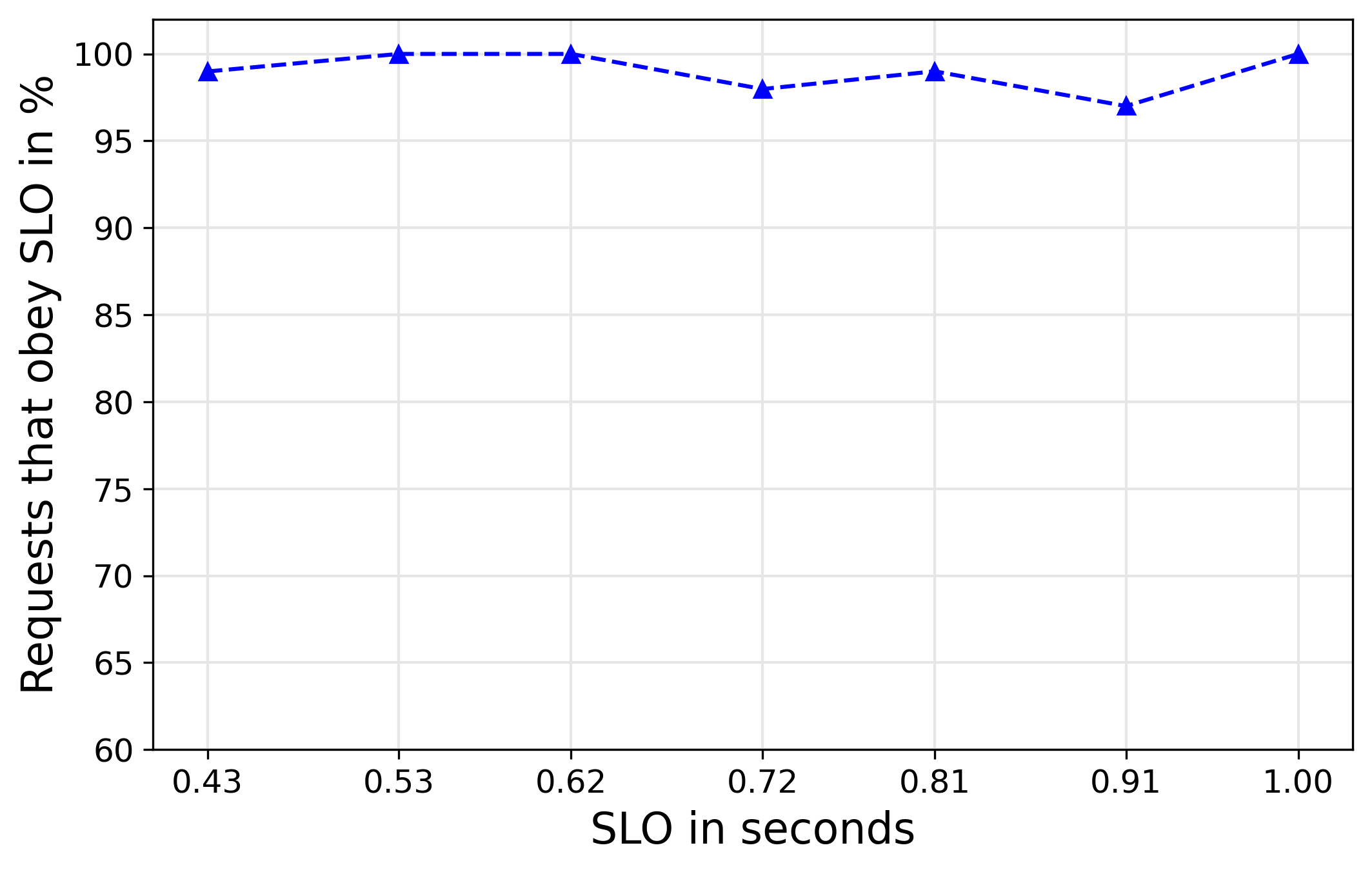}
    \caption{Real-world based application} \label{fig:RealPrercision} 
\end{subfigure}\hfill
\caption{
Percentage of the requests conforming to the given SLOs based on the configurations suggested by \textit{SLAM}.}
\label{fig:precision_applications}
\end{figure}

Here, we calculate the percentage of requests conforming to the defined SLOs when the functions are configured with the memory configurations suggested by \textit{SLAM-SLO} algorithm. Experiment results on the four test applications is shown in Figure~\ref{fig:precision_applications} for different SLOs when a total number of $100$ requests were issued to the application at each SLO. We can observe that for all the synthetic applications, the percentage of requests conforming to the given SLOs is either equal or above $95$\% which means that out of issued $100$ requests at least $95$ requests were served within the specified SLO execution time. Additionally, for the \textit{Real-world based} application as well, despite having lower estimation time accuracy as compared to synthetic applications, \textit{SLAMv} is still able to generate configurations that result in above $95$\% precision of requests conforming to the given SLOs. 

\subsubsection{Various objectives configuration finding effectiveness}
\label{sec:config_finding_effictiveness}
In this aspect, we determine the effectiveness of \textit{SLAM} tool when requested to optimize for various optimization objectives
(\S\ref{sec:objectives_optimization}) in addition to the SLOs. In this regard, we calculate the overall execution time and the cost needed by one invocation of the application when configured with memory configurations selected by \textit{SLAM} for those optimization objectives and compared them against static minimum-memory=$128$MB (\textit{min-mem}) and maximum-memory=$2$GB (\textit{max-mem}) configurations to get the worst/best execution times for the applications, and also the corresponding costs. The memory configurations of \textit{min-mem} and \textit{max-mem} signify configurations where each function in the application is configured to that memory. It is to be noted that, it is not necessary that we get the worst/best execution times at the extreme end of execution time~\cite{powerTuning}, therefore we compare them with the global minimum cost (\textit{BF-min-cost}) and execution time (\textit{BF-min-time}) for each application obtained by checking every configuration and function combinations using \textit{Brute force}. 

\begin{figure*}[t]
\centering
\begin{subfigure}{0.36\linewidth}
   \centering
    \includegraphics[width=1\linewidth]{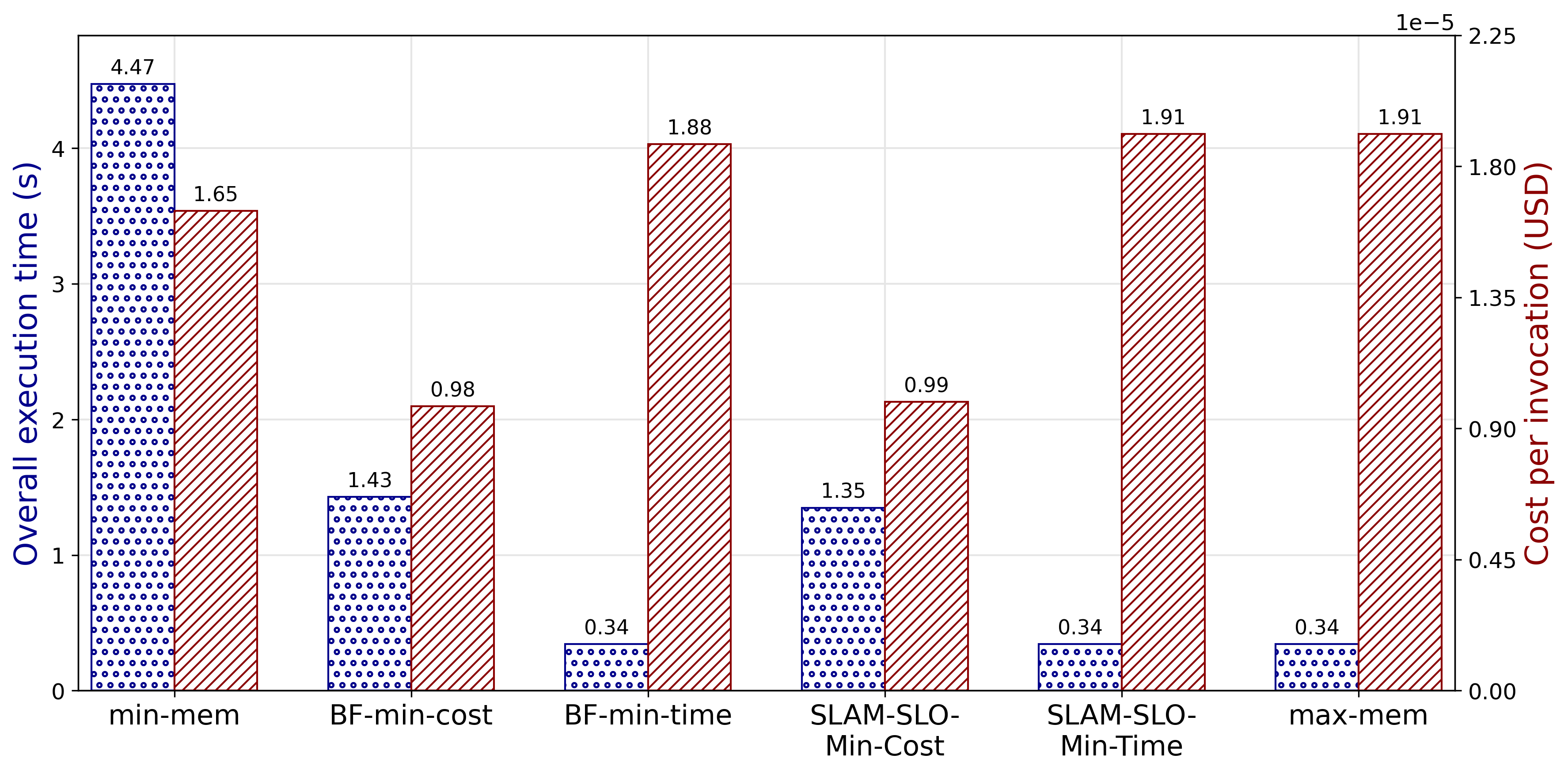}
    \caption{3-functions test application} \label{fig:3PriceDur}
\end{subfigure}
\begin{subfigure}{0.36\linewidth}
    \centering
    \includegraphics[width=1\linewidth]{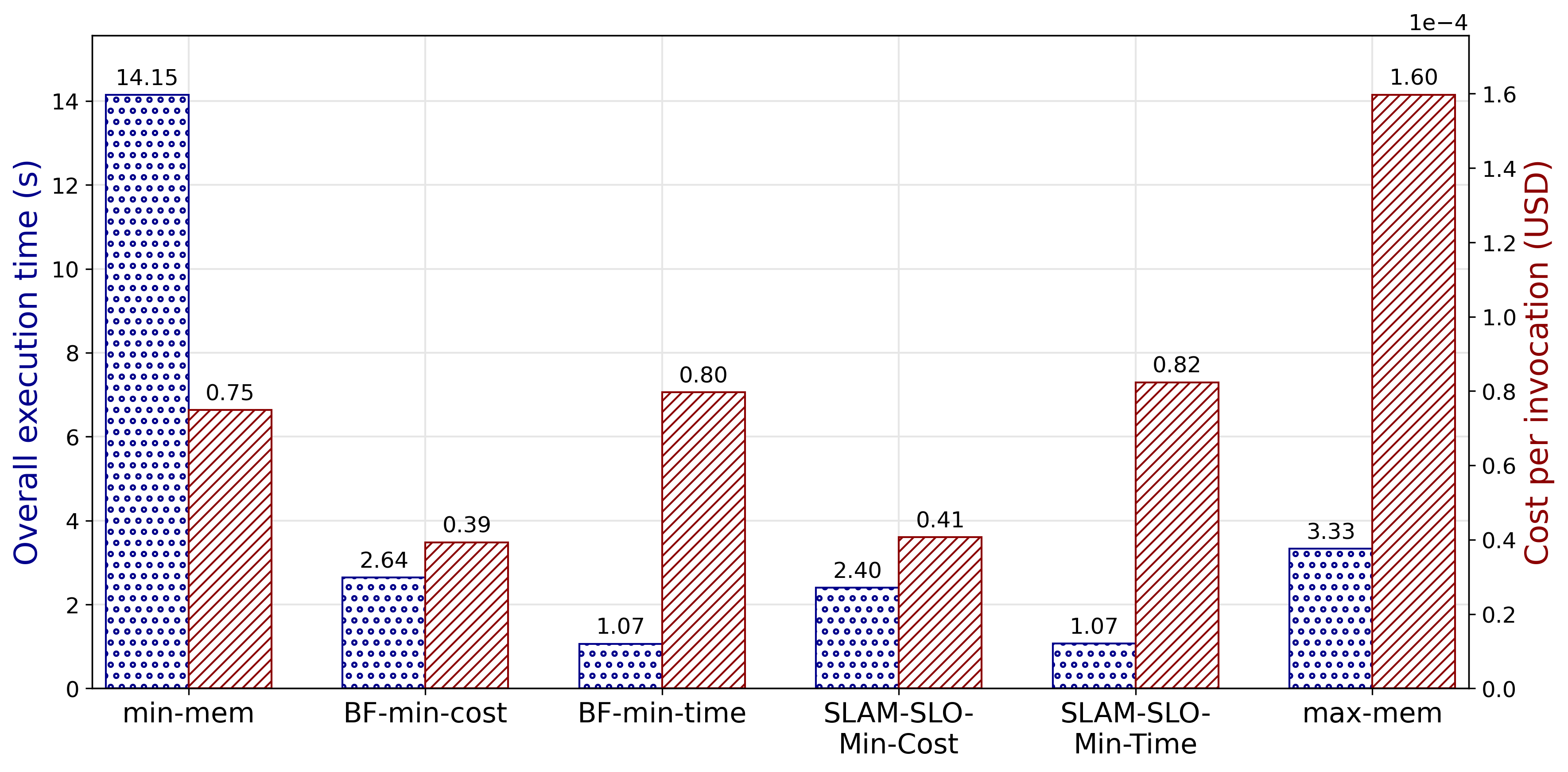}
    \caption{6-functions test application} \label{fig:6PriceDur}
\end{subfigure}
\begin{subfigure}{0.36\linewidth}
   \centering
    \includegraphics[width=1\linewidth]{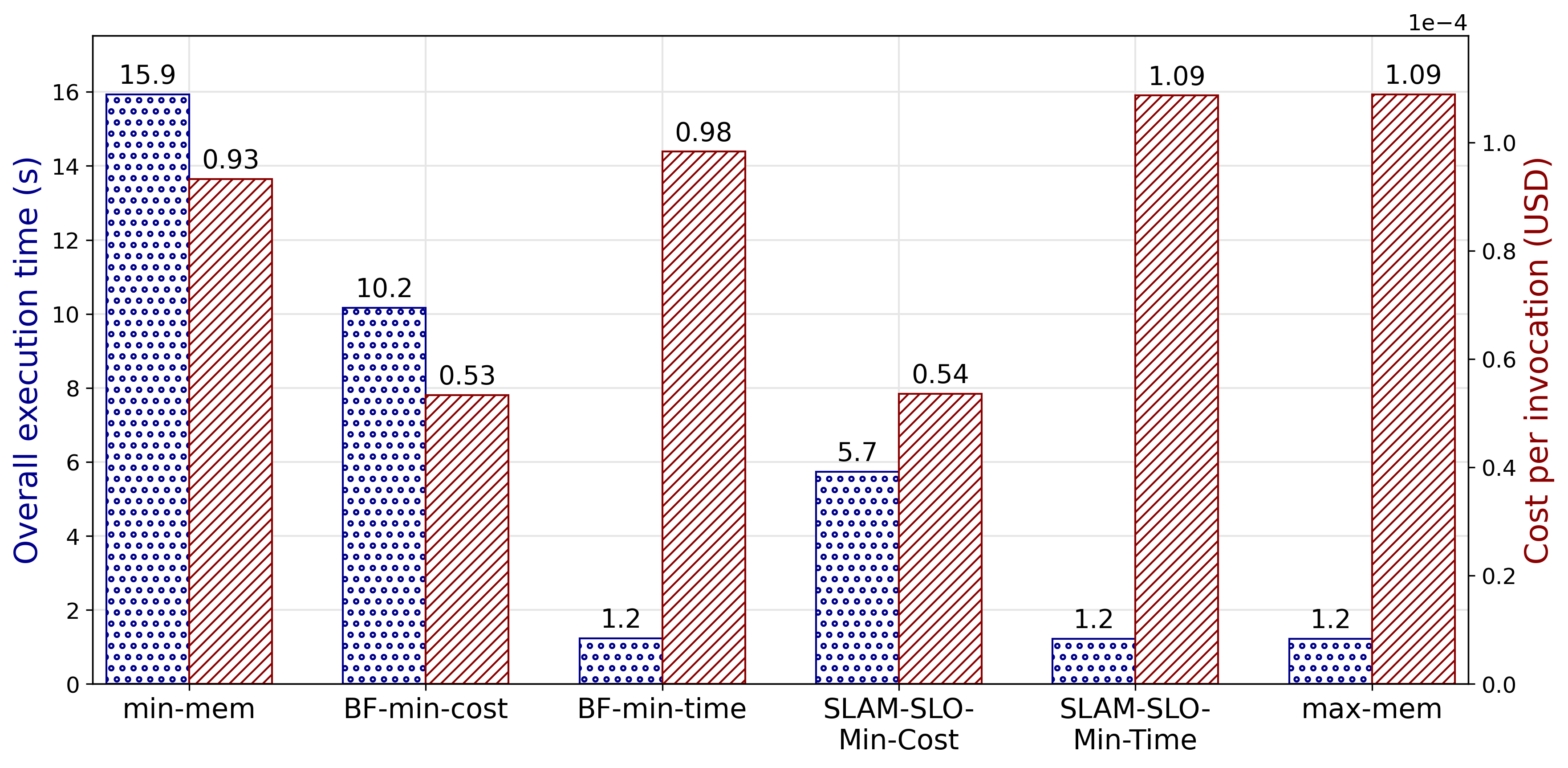}
    \caption{10-functions test application} \label{fig:10PriceDur} 
\end{subfigure}
\begin{subfigure}{0.36\linewidth}
   \centering
    \includegraphics[width=1\linewidth]{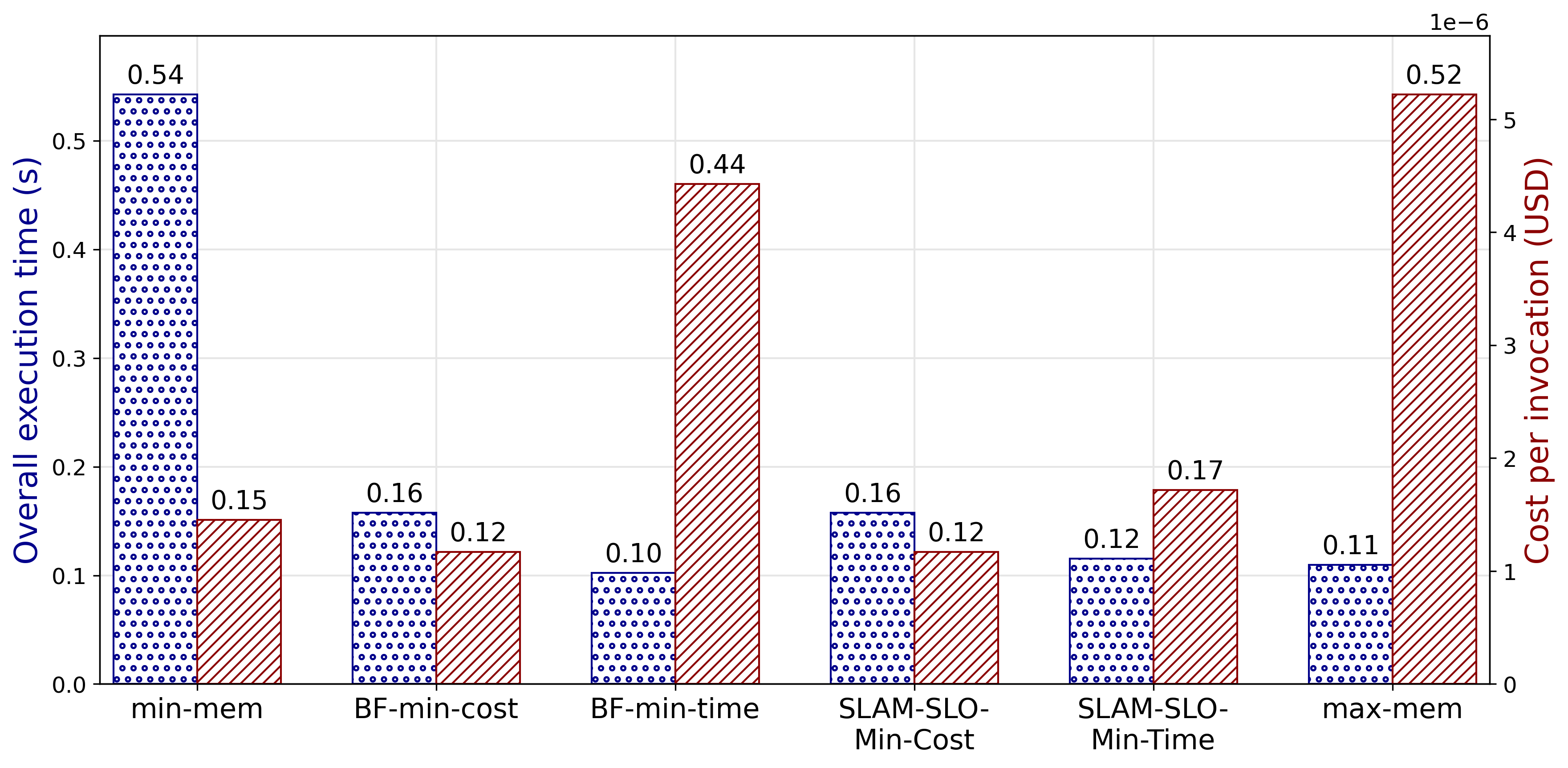}
    \caption{Real-world based application} \label{fig:RealPriceDur} 
\end{subfigure}
\caption{Execution time and the cost when configured with configurations selected by \textit{SLAM} for various objectives.}
\label{fig:price_dur_applications}
\end{figure*}

Experiment results on the four test applications are shown in Figure~\ref{fig:price_dur_applications} and the results are averaged over $100$ application invocations. From Figure~\ref{fig:price_dur_applications}, we can see that for all the applications, \textit{SLAM} optimization objective algorithms find the optimal/near-optimal cost and time configurations such that they are very close to the global minimum cost (\textit{BF-min-cost}) and time (\textit{BF-min-time}). Since the behavior of the \textit{SLAM} on different applications is very similar, we only explain the results for the \textit{3-functions} application on two objectives:

\textbf{Minimum Overall Cost}: For the \textit{3-functions} application, \textit{SLAM-SLO-Min-Cost} ($\$0.99 \times 10^{-5}$ as seen in Figure~\ref{fig:3PriceDur}) is only $\$0.01 \times 10^{-5}$  higher than \textit{BF-min-cost} ($\$0.98 \times 10^{-5}$).  When comparing \textit{SLAM-SLO-Min-Cost} with the \textit{min-mem} and \textit{max-mem} configuration, \textit{SLAM-SLO-Min-Cost} takes on average $1.6$x less cost than \textit{min-mem} and $1.9$x less cost than \textit{max-mem}.
Additionally, \textit{SLAM-SLO-Min-Cost} configuration ($1.3s$) is able to process application request faster than the \textit{min-mem} ($4.5s$) and \textit{BF-min-cost} configurations ($1.4s$)  but takes longer time than the \textit{max-mem} configuration ($0.3s$).

\textbf{Minimum Overall Execution Time}: 
For the \textit{3-functions} application, the execution time for \textit{SLAM-SLO-Min-Time} configuration ($1.07$s as seen in Figure~\ref{fig:3PriceDur}) is equivalent to that of \textit{BF-min-time} configuration and the overall cost for \textit{SLAM-SLO-Min-Time} ($\$0.82 \times 10^{-5}$) is only a bit higher than the \textit{BF-min-time} configuration ($\$0.80 \times 10^{-5}$). This shows that \textit{SLAM} is able to find the optimal/near-optimal execution time configuration such that it is very close to the global minimum execution time configuration (i.e.,  \textit{BF-min-time}, which requires a long time for determination).  From Figure~\ref{fig:3PriceDur} again we can see that the execution time taken by \textit{max-mem}
configuration ($3.3$s) is higher than that of \textit{BF-min-time} configuration ($1.07$s), therefore it may not always be true that the largest memory results in minimum overall execution time~\cite{powerTuning}. 
When comparing \textit{SLAM-SLO-Min-Time} configuration ($1.07$s) with the \textit{min-mem} ($14.15$s) and \textit{max-mem} ($3.3$s) configurations, \textit{SLAM-SLO-Min-Time} configuration takes on average $13.5$x less execution time than \textit{min-mem} configuration and $3$x less execution time than \textit{max-mem} configuration.  Additionally, \textit{SLAM-SLO-Min-Cost} configuration ($1.3s$) is able to process application request faster than the \textit{min-mem} ($4.5s$) and \textit{BF-min-cost} ($1.4s$) configurations but takes longer time than the \textit{max-mem} ($0.3s$) configuration. 
    
´

\subsection{Q3. \textit{SLAM} configuration finding efficiency and scalability}
\label{sec:config_finding_efficency_scalability}

\begin{figure}[t]
\centering
\begin{subfigure}{0.49\columnwidth}
   \centering
    \includegraphics[width=1\linewidth]{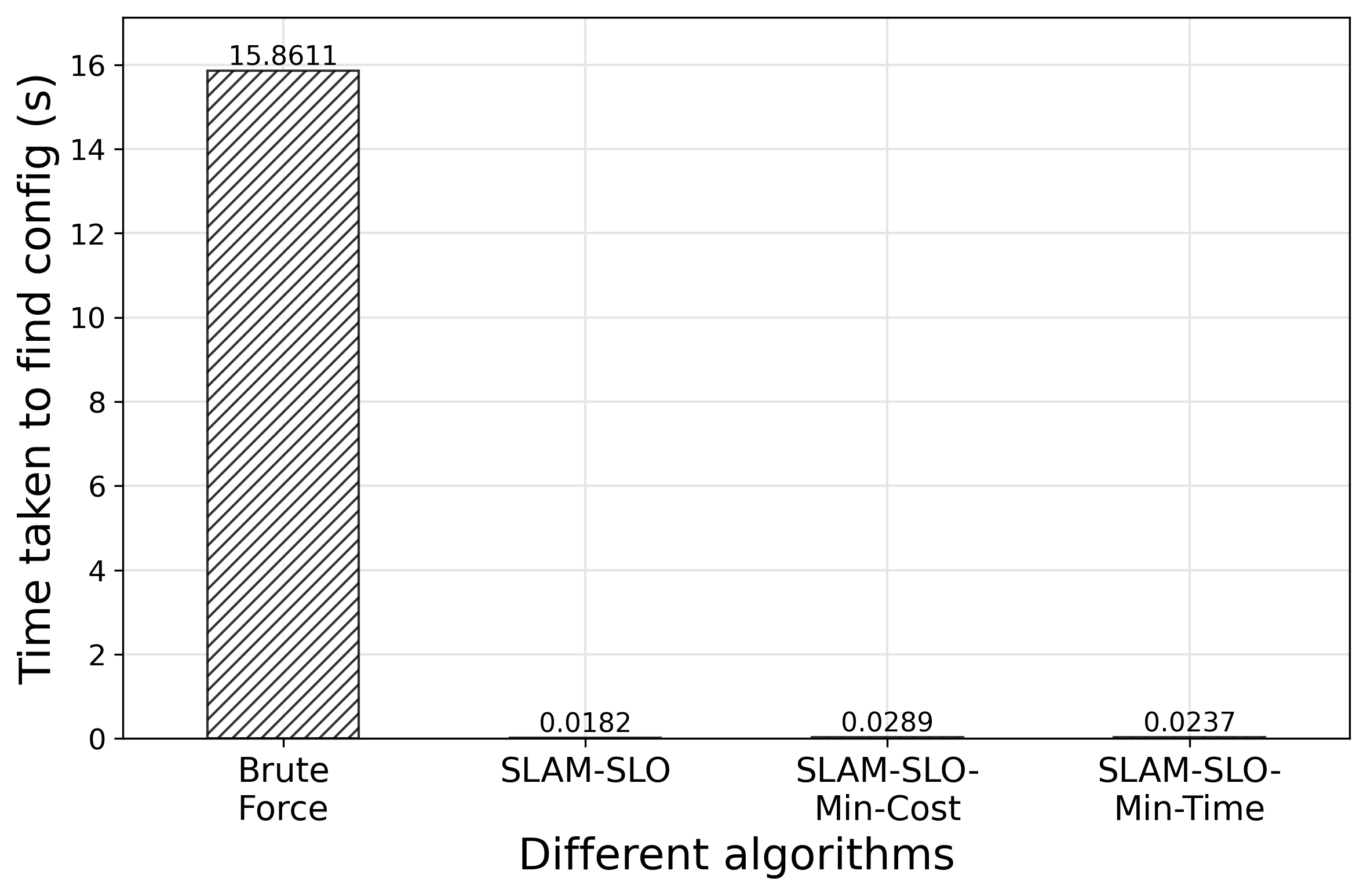}
    \caption{Configuration finding time for different algorithms} \label{fig:3DurCompAlgo}
\end{subfigure}\hfill
\begin{subfigure}{0.49\columnwidth}
    \centering
    \includegraphics[width=1\linewidth]{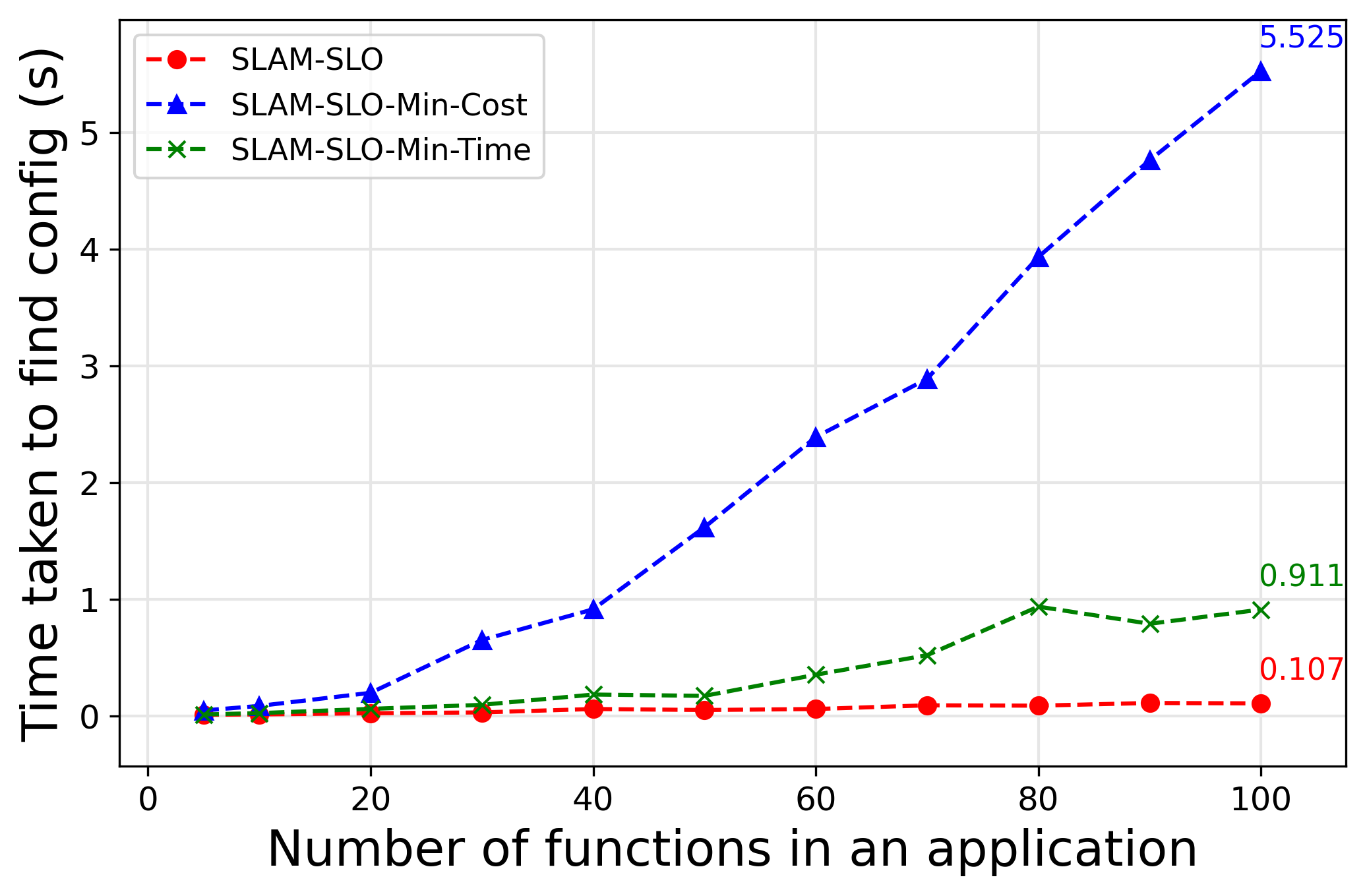}
    \caption{Performance of SLAM when the number of functions are scaled} \label{fig:scalability}
\end{subfigure}\hfill
\caption{SLAM efficiency and scalability performance }
\label{fig:scalability_efficiency}
\end{figure}

In Figure~\ref{fig:scalability_efficiency}, we show how efficient and scalable \textit{SLAM} is in finding the optimal configurations at various objectives. 
In Figure~\ref{fig:3DurCompAlgo} we can see the time required for different optimization algorithms to find the optimal configuration when run on \textit{6-functions} application. The \textit{Brute-force} algorithm performed worst as compared to the developed optimization algorithm (almost took $871$x time more than the developed algorithm). Although, it is possible to parallelize the \textit{Brute-force} search, but it is beyond the scope of this work.
When comparing \textit{SLAM-SLO} ($0.0182$s) with \textit{SLAM-SLO-Min-Cost} ($0.0289$s) and  \textit{SLAM-SLO-Min-Time} ($0.0237$s),  \textit{SLAM-SLO-Min-Cost} requires the most amount of time for this application with $6$ functions. This can also be validated from the Figure~\ref{fig:scalability} where the scalability of the three algorithms is tested on applications containing a larger number of functions (from $1$ to $100$) and  \textit{SLAM-SLO-Min-Cost} requires the most amount of time.  All algorithms scale linearly with the number of functions in the application, but with different slops and \textit{SLAM-SLO} having the least slope.

\textit{SLAM-SLO-Min-Cost}, which has to estimate the cost at every step of the search, has to go through a higher number of configurations as compared to \textit{SLAM-SLO} and \textit{SLAM-SLO-Min-Time}. Nevertheless, for an application containing $100$ functions \textit{SLAM-SLO-Min-Cost} took $5.5$s, which is not a lot considering the benefits of the algorithm in terms of cost-saving. 

\section{Related Work}

With the advent of serverless computing, there is a significant amount of research aimed at optimizing cloud computing resource utilization~\cite{Akin,Kulkarni, grogananalysis}. There has been some work on the performance profiling of various FaaS platforms.  Wang et al.~\cite{wang2018peeking} performed an in-depth study of resource management and performance isolation with three popular serverless computing providers: AWS Lambda, Azure Functions, and Google Cloud Functions. Their analysis demonstrates a reasonable difference in performance between the FaaS platforms. Furthermore, Shahrad et al.~\cite{shahrad2019architectural} studied the architectural implications of serverless computing and pointed out that exploitation of system architectural features like temporal locality and reuse are hampered by the short function runtimes in FaaS.  Chadha et al.~\cite{chadha2021architecturespecific} examine the underlying processor architectures for Google Cloud Functions (GCF) and determine the optimization of FaaS functions using Numba can improve performance by and save costs on average.  

Furthermore, there is a significant number of research works aimed at optimizing the memory and cost for the FaaS functions. COSE~\cite{cose} framework finds the optimal configurations for a FaaS function using the Bayesian Optimization algorithm while minimizing the total cost of execution. It not only models the behavior of a function, but also the environment (cloud, edge) in which those functions are deployed. However, they consider FaaS functions separately and optimized based on cost. Bayesian Optimization was also used in CherryPick~\cite{Cherrypick} tool for creating performance models for different cloud applications. The system provides $45$-$90$\% accuracy in finding optimal configurations and decreases cost up to $25$\%. But, they focused on traditional cloud applications.  Another framework Astra~\cite{astra}, is designed to optimize FaaS function configurations for specifically map-reduce usecase.

Similar optimization tools have also been developed by Google and Amazon. Google has developed a recommendation system to help the users choose the optimal virtual machine (VM) type~\cite{googleCloudReco}. It currently does not support Google Cloud Functions. AWS Compute Optimizer~\cite{awscomputeoptimzer} recommends optimal AWS resources for applications to reduce costs and improve performance by using machine learning to analyze historical utilization metrics. It can also be used to find optimal memory configuration for the lambda-based function. However, it can only be executed for the functions whose allocated memory level is less or equal to $1792$MB and which are invoked at least $50$ times in the last two weeks.  AWS Lambda Power Tuning~\cite{powerTuning} tool uses exhaustive search to identify optimal memory level for a cost, or execution time. By default, this algorithm will need to perform at least $225$ requests to the function to identify the optimal memory point.

None of the aforementioned research efforts address the issue of automatically configuring optimal memory of FaaS functions within a serverless application based on the user-defined SLOs. Most of the research either addresses a single FaaS function or an application consisting of step functions that do not have complex call graph workflows. The proposed tool \textit{SLAM} fills that gap by creating a recommendation tool that in a short time can find optimal memory configurations of FaaS functions within a serverless application given the SLOs. 

\section{Conclusion and Future work}

Serverless computing has abstracted most cloud server management and infrastructure scaling decisions away from the users, but configuring the memory of FaaS functions is still left up to the users. To solve this problem, we introduced  \textbf{SLAM} to find optimal memory configurations given predefined SLO requirements.
\textbf{SLAM} uses a max-heap-based optimization algorithm along with its variants for various optimization objectives (minimum cost and minimum overall time) in finding the optimal memory configuration for the given serverless application based on the specified SLO. It supports complex serverless application call-graph workflows and has the ability to adapt to changes in a serverless application. We demonstrate the functionality of \textit{SLAM} with AWS Lambda (\S\ref{sec:evaluation}) on four serverless applications consisting of a various number of functions and found that the suggested memory configurations guarantee that more than  95\%  of requests are completed within the defined  SLOs.  

In the future, we plan to extend \textit{SLAM} with other public serverless compute providers and to open source FaaS platforms. Transitioning from a discrete search space for the memory configurations to a continuous one could be the next improvement for the \textit{SLAM}. 

\bibliographystyle{IEEEtran}
\bibliography{bib}
\end{document}